\documentclass[journal,10pt]{IEEEtran}

\usepackage{amsmath}
\usepackage{suffix}
\usepackage{mathtools}
\usepackage{cite}
\usepackage{xcolor, soul}
\sethlcolor{green}
 
\usepackage{multirow}
\usepackage{hhline}

\usepackage[bottom]{footmisc}

\ifCLASSINFOpdf
\else
\fi

\usepackage{suffix,mathtools}
\DeclarePairedDelimiterX\MeijerM[3]{\lparen}{\rparen}%
{#3\,\delimsize\vert\begin{smallmatrix}#1 \\ #2\end{smallmatrix}}

\newcommand\MeijerG[8][]{%
  G^{\,#2,#3}_{#4,#5}\MeijerM[#1]{#6}{#7}{#8}}

\WithSuffix\newcommand\MeijerG*[7]{%
  G^{\,#1,#2}_{#3,#4}\MeijerM*{#5}{#6}{#7}}

\begin{document}

\title{Mixed RF-VLC Relaying Systems for Interference-Sensitive Mobile Applications}

\author{Milica~I.~Petkovic,~\IEEEmembership{Member,~IEEE,}~Aleksandra~M.~Cvetkovic,~\IEEEmembership{Member,~IEEE,}\\
Milan~Narandzic,~\IEEEmembership{Member,~IEEE,}~Nestor~D.~Chatzidiamantis,~\IEEEmembership{Member,~IEEE,}\\
Dejan~Vukobratovic,~\IEEEmembership{Senior  Member,~IEEE,}~and~George~K.~Karagiannidis,~\IEEEmembership{Fellow,~IEEE}

\thanks{M.~I.~Petkovic, M. Narandzic and D. Vukobratovic are with University of Novi Sad, Faculty of Technical Science, Novi Sad, Serbia (e-mails: milica.petkovic@uns.ac.rs; orange@uns.ac.rs; dejanv@uns.ac.rs).}
\thanks{A.~M.~Cvetkovic is with University of Nis, Faculty of Mechanical Engineering, Nis, Serbia (e-mail: aleksandra.cvetkovic@elfak.ni.ac.rs).}
\thanks{N.~D.~Chatzidiamantis and G.~K.~Karagiannidis are with the Aristotle University of Thessaloniki, Thessaloniki 54 124, Greece (e-mails: nestoras@auth.gr; geokarag@auth.gr).}
}

\maketitle

\begin{abstract}

Due to their Radio-Frequency (RF) immunity, Visible Light Communications (VLC) pose as a promising technology for interference sensitive applications such as medical data networks. In this paper, we investigate mixed RF-VLC relaying systems especially suited for this type of applications that support mobility. In this system setup, the end-user, who is assumed to be on a vehicle that is in dynamic movement, is served by an indoor VLC system, while the outdoor data traffic is conveyed through multiple backhaul RF links. Furthermore, it is assumed that a single backhaul RF link is activated by the mobile relay and due to feedback delay, the RF link activation is based on outdated channel state information (CSI). The performance of this system is analyzed in terms of outage probability and bit error rate (BER), and novel closed form analytical expressions are provided. 
Furthermore, the analysis is extended for the case where the average SNR over the RF links and/or LED optical power is high, and approximate analytical expressions are derived which determine performance floors. Numerical results are provided which demonstrate that the utilization of multiple RF backhaul links can significantly improve overall RF-VLC system performance when outage/BER floors are avoided. This calls upon joint design of both subsystems. Additionally, the outdated CSI exploited for active RF selection can significantly degrade the quality of system performance. 
\end{abstract}


\IEEEpeerreviewmaketitle

\section{Introduction}
With the constant increase in the number of users, the development of modern technologies and adaptation of the existing ones is mandatory in next generation of wireless communication systems.  The challenge to respond to novel  demands, such as higher data rates, improved security and wider coverage area, leads to the intensive  research and industrial interests in novel communication technologies.
As an innovative  modern technology for indoor and outdoor applications, the optical wireless communications (OWC) have received  attention in  research and industry areas,  offering a number of benefits, such as large bandwidth, support for more users, license-free operation, low-cost \cite{OWC_MATLAB,bookvlc, ComMag, vlc1, Haas}. Of particular interest are the indoor OWC systems, known as visible light communications (VLC) that operate at the optical wavelengths of the 380-750 nm
which belong to visible spectrum and can be particularly attractive for "interference-sensitive" applications, i.e., applications which have increased throughput requirements and are critical not to cause interference or be interfered by other radio-frequency (RF) systems \cite{OWC_MATLAB,bookvlc, ComMag,new3}. Such applications are usually encountered in medical health care data systems as presented in  \cite{Ray1}. 
It is worth mentioning that new concepts such as smart VLC integrate both illumination and communication functionality \cite{smartVLC}.
 
\begin{figure*}[!t]
\normalsize

\centering
\includegraphics[width=17cm,height=6.3cm]{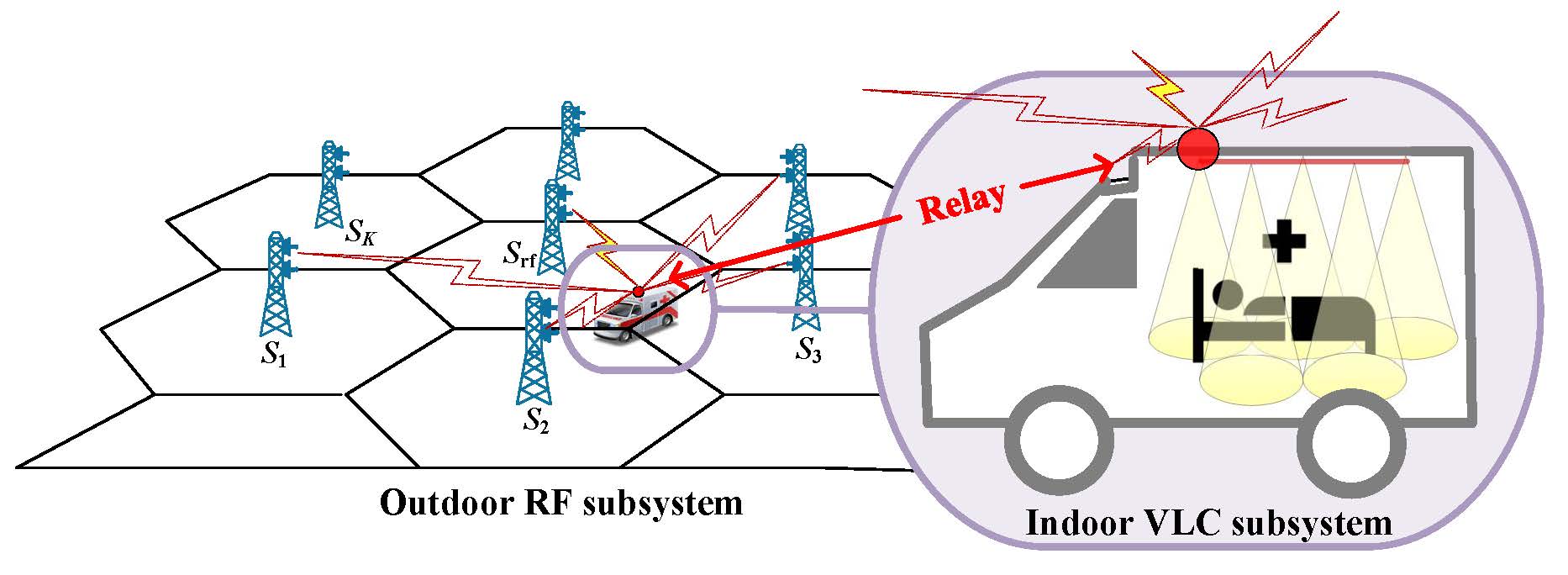}
\caption{System model of a dual-hop mixed RF-VLC communication system for interference sensitive applications.}
\label{Fig1}

\vspace*{6pt}
\end{figure*}

According to listed benefits, the OWC systems represent an appropriate alternative or complement to the traditional RF signal transmission. Due to the widespread installation of RF communication systems, their combinations with indoor OWC systems are easily envisioned. In the resulting topology of heterogeneous system, it is possible to distinguish between parallel and serial aggregation of communication links (combining radio and/or optical transmission). The serial concatenation of links, which is in the focus of this paper, is typically referred to as mixed RF-VLC relaying. Since relaying ensures wider coverage area and/or improved data rate capacity, its employment is commonly considered for the 
forthcoming communication systems.

\subsection{Literature overview}
In the past, various combinations of the RF and VLC links as heterogeneous systems have been investigated \cite{hybridRFVLC,RF1,RF3,RF4}. Specifically, hybrid RF/VLC downlink system based on hard switching implemented in indoor environment was investigated in \cite{hybridRFVLC}.
Integration of VLC and RF network was discussed in terms of the coverage/rate analysis and energy efficiency  in \cite{RF1}, while the energy efficiency perspective was considered in \cite{RF3}.  Secrecy outage probability of the hybrid  VLC/RF system where the legitimate receiver harvests energy from the LED based on stochastic geometry theory, was studied in \cite{RF4}.

The utilization of the relaying technology within mixed RF-VLC relaying systems was considered in \cite{EH1,EH3,Ray1,Ray2,Ray4,konf, VLC-RF-new}. Specifically, dual-hop VLC-RF  systems with the relay harvesting energy from the received optical signal was studied in \cite{EH1,EH3}. 
Performance of the RF-VLC relaying  system  with decode-and-forward (DF) relay was analysed  in \cite{Ray2,konf},  while  the outage probability of the VLC-RF relaying system with multiple DF relays was derived in \cite{Ray4}. Furthermore, the outage probability  performance of the RF-VLC system with one amplify-and-forward (AF) relay  was studied in \cite{Ray1}. Finally, outage probability and error rate performance of VLC-RF   relaying  system was analysed  in \cite{VLC-RF-new}, considering that DF or AF  relay  is randomly positioned in the coverage area of the LED lamp. 

\subsection{Motivation and contribution}
Motivated by aforementioned works, in this paper, we aim at designing  mixed RF-VLC relaying systems suited for interference sensitive mobile applications. In these application scenarios, the end-user is assumed to be on a vehicle which is in dynamic movement (e.g. emergency ambulance, trains, airplanes or  upcoming self-driving electrical vehicles), while the indoor environment is RF unfriendly \cite{Ray1}, i.e., strong electric field intensity induced by some RF frequencies can interfere with electronic equipment resulting in critical data loss \cite{new3}. In this type of systems, multiple backhaul RF links are utilized to convey the outdoor traffic, while  the  VLC system is used to deliver the  data to the end mobile user in indoor environment. Differently to \cite{model}, which considered Free Space Optical (FSO) link as a backhaul link solution, utilization of the RF links for outdoor traffic is inspired by the fact that the required line-of-sight (LoS) condition for FSO transmission is difficult to accomplish, especially in urban mobile environment. As depicted in Fig.~\ref{Fig1}, broadband service is provided to the end user  by the indoor VLC access point with support of the multiple backhaul RF links.  

Based on the above, the contribution of the paper is summarized as follows:
\begin{enumerate}
\item We design a RF-VLC relaying system specially suited for interference sensitive mobile applications. In particular, multiple base stations (BSs) are assumed to offer high-capacity backhaul options for the indoor VLC access point, and one serving BS, i.e., "backhaul" RF link, is selected by relay. It should be noted that this operation mode is equivalent to mobile evaluated handover mode, where user makes decision about targeted cell \cite{meho1} and radio-access diversity is established by selecting the RF backhaul link as the best one among all possible  RF links. However, due to feedback delay, we assume that active BS selection is based on \textit{outdated} channel state information (CSI).
\item An analytical framework for the performance evaluation of the mixed  RF-VLC relaying system under consideration is provided. Both fixed gain AF and DF relaying schemes are taken into consideration during the  performance analysis.
 Specifically, assuming that RF links are subject to Nakagami-\textit{m} fading (which efficiently models both LoS and non-LoS transmissions) and VLC link is subject to geometry-dependent channel model \cite{OWC_MATLAB, Haas, model}, we derive analytical closed-form expressions for the outage probability and the average bit error rate (BER). Furthermore, we extend the analysis to include the cases of high average SNR over RF link and large LED optical power, and provide approximate analytical expressions, which determine outage probability and average BER  floors of the considered system. Moreover, the approximate expressions corresponding to the both high average SNR over RF link and large LED optical power are also derived.
\item  Numerical and simulation results are provided to verify the presented analysis and illustrate the effects of channel and system parameters on the system performance. 
\end{enumerate}

\subsection{Structure}

The remainder of the paper is organized as follows. Section II describes the RF-VLC relaying system model under consideration, while the  channel model of both RF and VLC links is given in Section III. Furthermore, the analytical results regarding the outage probability and the average BER analysis of the system under consideration are provided in Sections IV and V, respectively. Numerical results illustrating the effects of channel parameters on system performance are depicted at Section VI. Finally, Section VII concludes the paper.

\section{System  model}

As presented in Fig.~\ref{Fig1}, proposed dual-hop  RF-VLC  system includes  
$ K $ BSs, denoted by $ S_k $, $ k=1,\ldots,K $, a   relay node, denoted by $ R $, and a mobile end user located in indoor environment. 
The $k$-th BS, $S_k$, for $ k=1,\ldots,K $, can transmit an electrical signal, denoted by $x_k$,  with the average transmitted electrical power  $P_s$, to an AF or a DF relay via RF link. The signal received from the $k$-th BS at the relay node can be determined as
\begin{equation}
y_k = h_kx_k + n_r,
\label{rR}
\end{equation}
where  $h_k$ denotes the fading amplitudes of the $S_k - R$ link with an average power normalized to one, i.e., ${\rm E}\left[ |h_k|^2 \right] = 1$  where $ {\rm E}\left[ \cdot \right]$ denotes mathematical expectation, and $n_r$ is the complex additive white Gaussian noise (AWGN) with zero mean and variance $\sigma_r^2$  at the relay. The instantaneous signal-to-noise ratio (SNR) at the relay is defined as
\begin{equation}
\gamma_{k} = \frac{|h_k|^2P_s}{\sigma_r^2}= |h_k|^2 \mu_{\rm rf},
\label{snrRF}
\end{equation}
with the average SNR  determined as
\begin{equation}
\mu_{\rm rf} = {\rm E}\left[ \gamma_{k} \right] = \frac{{{\rm E}\left[|h_k|^2 \right] P_s}}{\sigma _r^2} = \frac{P_s}{\sigma _r^2},
\label{mi1}
\end{equation}
with assumption that the average SNRs for all $k=1, \cdots, K$ RF links are equal.

The RF signal transmission  is performed over the single active RF link - the one with the best estimated channel condition (highest fading amplitude $h_k$ and instantaneous SNR)  among all $S_k - R$ links. For properly designed cellular system, the co-channel interference (the same channel repeated in nearby interfering cell) should be far below signal level, so it is omitted from our analysis \cite{R1}.
Due to  feedback delay, time channel variations occur, thus estimated CSI  used for selection of the active BS happens to be time-delayed, i.e., outdated. The  outdated version of $\gamma_k$ used for channel estimation, denoted by $\tilde \gamma_k$, in general differs from actual instantaneous SNR value. The similarity between outdated and actual value of $\gamma_k$ is expressed by correlation coefficient $\rho$.
Consequently,  an estimation error occurs and the active RF link is not necessarily the best one among the set of all $S_k - R$ links.

Hence, the selected BS is determined by
\begin{equation}
\begin{split}
 n = {\rm arg~max}_k \left\{  \tilde \gamma_{k} : k=1, \cdots , K \right\},
\label{n_rf}
\end{split}
\end{equation}
while the instantaneous SNR of the active  RF link is determined as  
\begin{equation}
\begin{split}
\gamma_{\rm rf} = \gamma_n = \frac{|h_n|^2 P_s}{\sigma_r^2}= |h_n|^2 \mu_{\rm rf},
\label{snr_rf}
\end{split}
\end{equation}
where $h_n$ represents the  fading amplitude of the selected   RF link.

\subsection{AF relay}
For high transmission power for RF link, the fixed gain AF relay is implemented, thus the  amplification is performed based on long-term statistics of the RF channel, i.e., the relay gain $ G $ can be determined as \cite{gain}
\begin{equation}
G^2  = \frac{\xi}{\sigma_{r}^2 \left( {\rm E}\left[ \gamma _{\rm rf} \right] + 1 \right)}= \frac{\xi}{\sigma _{r}^2 C  },
\label{gainG}
\end{equation}
where a fixed gain constant is defined as $ C = {\rm E}\left[ \gamma _{\rm rf} \right] + 1 $ and $\xi$  is the constant related to optimal power level adjustment from RF to VLC (in downlink) that accounts for the conversion of electrical signals (taking both positive and negative values) into optical (only positive) signals. Proper $\xi$ value avoids signal clipping and takes care of power constraints. Without loss of generality we assume that $\xi=1$\footnote{Please note that the power level adjustment scales all appearances of $C$ with constant $1/\xi$.}, thus $G^2  = \frac{1}{\sigma _{r}^2 C  }$. Afterwards, intensity modulation is performed by adding a DC bias, which is removed at the receiver.

The second hop is assumed to be in indoor surrounding, containing multiple LED lamps placed on the ceiling. The  mobile receiver terminals are uniformly distributed over the coverage area of the room.  
The mobile user receives the optical signal from a VLC access point that provides the most powerful
channel gain, while the signals from other LED lamps, i.e.,  intercell interference, are treated as a Gaussian noise \footnote{We assume that  each LED  lamp is characterized by an unique random ID sequence to encode the information bits. The user terminals have a knowledge about the ID sequences. After optical-to-electrical signal conversion, the user can compare the signals received from all LEDs and determine the strongest one based on these ID sequences. In this way, the user knows which LED lamp provides the strongest signal, and selects it for further information processing.}   \cite{ISI}. 
This means that  the best link is selected in both outdoor and indoor environments. 
After this point, the system model is simplified to one BS in outdoor environment and one LED inside the vehicle. At the destination, direct detection and optical-to-electrical signal  conversion is done via PIN photodetector with the conversion efficiency denoted by $\eta$.
Finally, the electrical signal at the  destination is given by 
\begin{equation}
s = P_t I \eta G y_{\rm rf} + n_d = P_t  I\eta G \left( h_n x_n + n_r \right) + n_d,
\label{rD}
\end{equation}
where $y_{\rm rf} = h_n x_n + n_r$  is received electrical signal at the relay node from the active  BS,  $P_t$ is the average transmitted optical power of a LED lamp,   $I$  represents the  DC channel gain of the LoS link between LED lamp and the end user,  and $n_d$ is the AWGN over VLC link with zero mean and variance $\sigma_d^2=N_0W$, where $N_0$ denotes noise spectral density and $W$ is the baseband modulation bandwidth.  Furthermore, for the purpose of the analysis, it is adopted that the lenses are employed as a part of the LED lamp to regulate
direction and focus of the LED lighting.  Although the VLC channels include both LoS and diffuse components, 
the reflected signals energy is neglected in the further system performance analysis  since it is significantly lower than the energy of the LoS component  \cite{OWC_MATLAB, LOS}.

The overall end-to-end SNR at the destination can  be determined based on (\ref{snr_rf}), (\ref{gainG}) and (\ref{rD}) as
\begin{equation}
\gamma^{\rm (af)} _{\rm eq} = \frac{P_t^2  I^2 \eta^2 G ^2h_n^2 P_s }{P_t^2  I^2 \eta^2 G ^2\sigma_r^2 + \sigma_d^2} = \frac{\gamma_{\rm rf} \gamma_{\rm vlc}}{\gamma_{\rm vlc} +  C}.
\label{endSNR}
\end{equation}
where the instantaneous SNR of the VLC link is defined as 
\begin{equation}
\gamma_{\rm vlc} = \frac{P_t^2 I^2 \eta ^2}{\sigma _d^2}.
\label{snrVLC}
\end{equation}

\subsection{DF relay}

When  DF based RF-VLC relaying system is considered, the instantaneous equivalent end-to-end SNR, $ \gamma^{\rm (df)}_{\rm eq} $, can be defined as
\begin{equation}
\gamma^{\rm (df)}_{\rm eq} = \min \left( \gamma_{\rm rf},\gamma _{\rm vlc} \right),
\label{endSNRdf}
\end{equation}
where $\gamma _{\rm rf}$  is the instantaneous SNR defined in (\ref{snr_rf})  and $\gamma _{\rm vlc}$   is the instantaneous SNR of the  VLC link defined in (\ref{snrVLC}).

\section{Channel model}

\subsection{RF channel model}

Since  RF links experience independent and identically distributed  Nakagami-\textit{m} fading\footnote{Assumption of identical small-scale fading distributions is supported by power control mechanism of the cellular network and future network densification that mitigate effects of the transmission loss and large-scale fading (shadowing) on moving terminal.},  by arranging  $\tilde  \gamma_{k}$ for $k~=~1,\cdots,K$ in an increasing order of magnitude by using the aproach described in Appendix A,   the probability density function (PDF)  and cumulative density function (CDF) of the instantaneous SNR of the active RF link, $\gamma_{\rm rf}$, are respectively determined as 
\begin{equation}
\begin{split}
 f_{\gamma _{\rm rf}}\left( \gamma \right)& = K\sum\limits_{p = 0}^{K- 1}\sum\limits_{\bigtriangleup = p}  \sum\limits_{i = 0}^B A \frac{\rho ^i{\left( {1 - \rho } \right)}^{B - i}}{{{{\left( {1 + p\left( {1 - \rho } \right) } \right)}^B} }}\\
&\times  {\left( {\frac{Q}{{\left( {1 + p} \right)}}} \right)^{m + i}}{e^{ - Q\gamma}}{\gamma^{m + i - 1}},
\label{pdf_rf}
\end{split}
\end{equation}
\begin{equation}
\begin{split}
 F_{\gamma _{\rm rf}}\left( \gamma \right)&= 1- K\sum\limits_{p = 0}^{K - 1}\sum\limits_{\bigtriangleup = p}  \sum\limits_{i = 0}^B A \frac{\rho ^i{\left( {1 - \rho } \right)}^{B - i}}{{{\left( 1 + p\left( {1 - \rho } \right) \right)}^B}}\\
& \times \frac{{\Gamma \left( {m + i,Q\gamma } \right)}}{{{{\left( {1 + p} \right)}^{m + i}}}},
\label{cdf_rf}
\end{split}
\end{equation}
where fading severity parameter is expressed by $m$ and $ \Gamma \left( \cdot, \cdot \right)$  denotes the Incomplete Gamma function defined in \cite[(8.350.2)]{grad}. The second sum in (\ref{pdf_rf}) and (\ref{cdf_rf}) 
\begin{equation}
\bigtriangleup =  \sum\limits_{i = 0}^{{m} - 1}  {{l_i} },
\label{constdelta}
\end{equation}
contains all \textit{m}-tuples $(l_0, \cdots , l_{m-1})$ of nonnegative integers whose sum is $p$. Furthermore,
\begin{equation}
B = B\left( {{l_0},{l_1},...,{l_{m - 1}}} \right) = \sum\limits_{j = 0}^{m - 1} {j\,{l_j}} ,
\label{constB}
\end{equation}
\begin{equation}
\begin{split}
 A &= A\left( {p,i,{l_0},{l_1},...,{l_{m - 1}}} \right) \\
&=\left( - 1\right)^p {K-1 \choose p}{p \choose {{l_0},{l_1},...,{l_{m - 1}}}} \\
& \times \frac{{\Gamma \left( {m + B} \right)\Gamma \left( {B + 1} \right)}}{{i!\Gamma \left( m \right)\Gamma \left( {m + i} \right)\Gamma \left( {B - i + 1} \right)}}\prod\limits_{j = 0}^{m - 1} {{{\left( {\frac{1}{{j!}}} \right)}^{{l_j}}}},
\label{constA}
\end{split}
\end{equation}
and
\begin{equation}
Q = Q\left( p \right) = \frac{{m \left( {1 + p} \right)}}{{\left( {1 + p\left( {1 - \rho } \right)} \right){\mu _{\rm rf}}}}.
\label{const2}
\end{equation}

The fixed gain constant $ C = {\rm E}\left[ \gamma _{\rm rf} \right] + 1 $ in (\ref{gainG}) can be  determined based on (\ref{pdf_rf})  as 
\begin{equation}
\begin{split}
 C\!  =\!  1\! +\!  K\sum\limits_{p = 0}^{K - 1} \! \sum\limits_{\bigtriangleup = p} \sum\limits_{i = 0}^B  \frac{A\rho ^i{\left( {1 - \rho } \right)}^{B - i}{\Gamma \left( {m + i + 1} \right)}}{{{\left( 1 + p\left( {1 - \rho } \right) \right)}^B {{{\left( {1 + p} \right)}^{m + i}}Q}}}.
\label{const}
\end{split}
\end{equation}

\subsection{VLC channel model}

\begin{figure}[!b]
\centering
\includegraphics[width=2.5in]{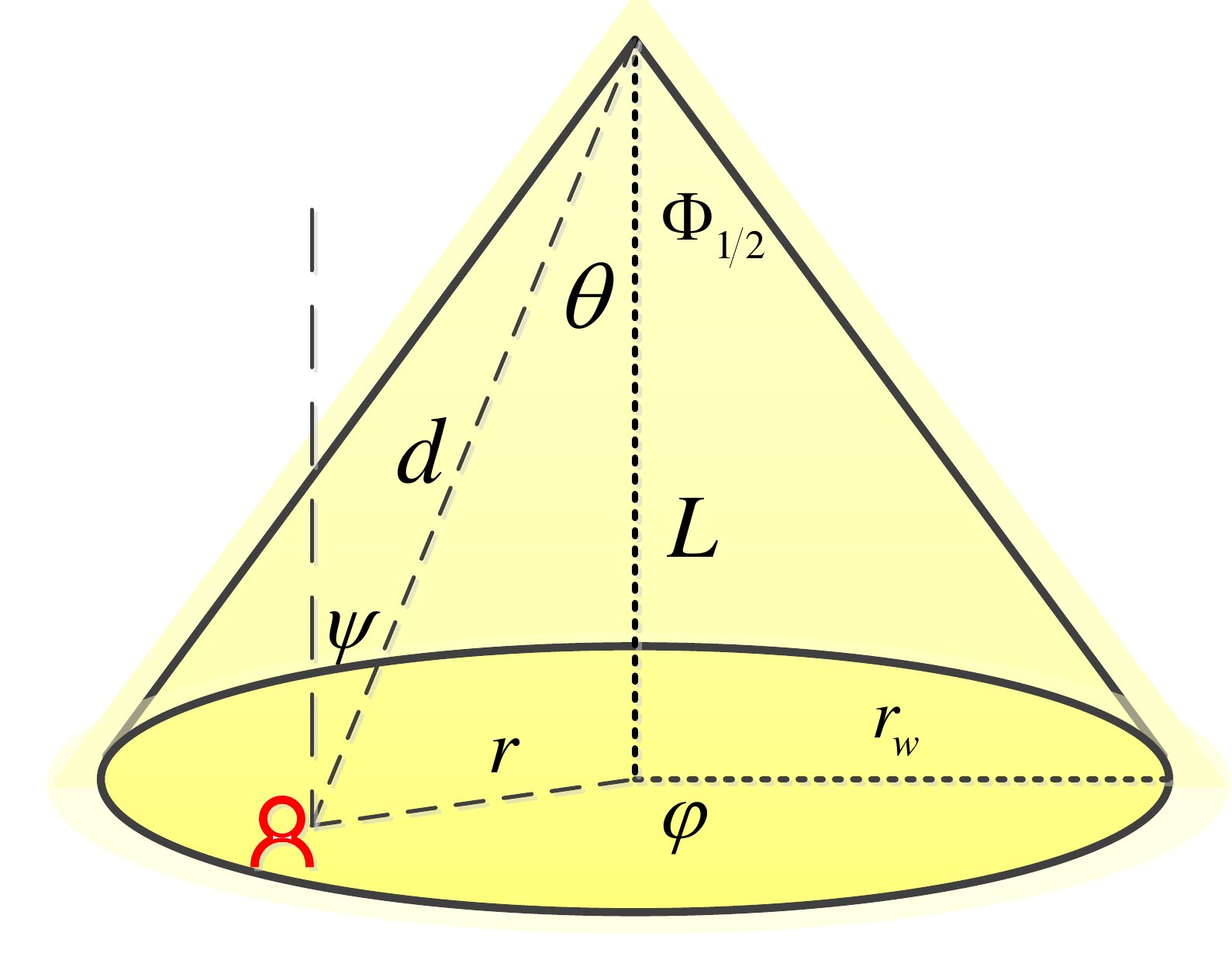}
\caption{Geometry of the LoS VLC propagation model.}
\label{Fig2}
\end{figure}

Related to indoor VLC subsystem, the LED  transmitter is positioned at height $ L $  from the receiving plane, as it can be observed in Fig.~\ref{Fig2}. The location of the end user is determined with angle of irradiance $\theta$, and  angle $\varphi$  and radius $r$  in the polar coordinate plane.  Moreover,   $\psi$ represents the angle of incidence, while the Euclidean distance between the  LED lamp and the photodetector receiver is denoted by $ d $. 
 
The DC channel gain of the LoS link between LED lamp and the mobile end user  is determined as \cite{OWC_MATLAB}
\begin{equation}
 I = \frac{\mathcal A\left( w + 1\right)\mathcal R}{2\pi d^2}\cos ^w\left( \theta \right)Tg\cos \left( \psi \right),
\label{I_n1}
\end{equation}
where $ \mathcal A $ is a physical surface area of photodetector receiver, $\mathcal R$ represents the responsivity, and $T$ is the gain of the optical filter. The optical concentrator is defined as $g =\zeta^2 /\sin^2\left( \Psi \right)$, for $0\leq\psi \leq \Psi$, where $\zeta$ is the refractive index of lens at a photodetector, and $\Psi$ denotes the field of view (FOV) of the receiver. The LED transmission  is assumed to follow a generalized Lambertian radiation pattern with the order $w$, which is related to the semi-angle at the half illuminance of LED, denoted by $\Phi_{1/2}$, as $w = -\ln 2/ \ln \left( \cos \Phi_{1/2} \right)$  \cite{OWC_MATLAB}.
Next, the  semi-angle at the half illuminance of LED is related by the maximum radius of a LED lighting footprint, $r_{w}$, as   $r_{w} =  L\tan \left(\Phi_{1/2} \right) $.
If the surface of photodetector receiver is parallel to the ground plane and has no orientation towards the LED, then $\theta =\psi $,   $d~=~\sqrt {r^2 + L^2} $,    $\cos \left( \theta \right) \!=\! \frac{L}{ \sqrt {r^2 + L^2} }$, and (\ref{I_n1}) can be rewritten as
\begin{equation}
I = \frac{\mathcal X  }{\left( r^2 + L^2 \right)^{\frac{w + 3}{2}}},
\label{I_n2}
\end{equation}
where  $\mathcal X = \frac{\mathcal A\left( w + 1 \right)\Re}{2\pi}Tg\left( \psi \right)L^{w + 1}$.
The mobile end user is assumed to be positioned within circular area covered by LED lighting. 
If the position of the end user is  modeled by a uniform distribution,  the PDF of the  radial distance is \cite{model}
\begin{equation}
f_{r}\left( r \right) = \frac{2r}{r_{w}^2},\quad 0\leq r \leq r_{w}.
\label{pdf_rn}
\end{equation}
Based on (\ref{I_n2}) and (\ref{pdf_rn}), after utilization of the  technique for transformation of random variables, the PDF of the channel gain, $  I $, is derived as \cite{model}
\begin{equation}
f_{ I}\left(  I \right) = \frac{2 \mathcal X ^{\frac{2}{w + 3}}}{r_{w}^2\left( w + 3 \right)}  I^{ - \frac{w + 5}{w + 3}},\quad  I_{\min }\leq  I \leq  I_{\max },
\label{pdf_In}
\end{equation}
where $ I_{\min} = \frac{\mathcal X }{{\left( r_{w}^2 + L^2 \right)}^{\frac{w + 3}{2}}}$ and  $I_{\max} = \frac{\mathcal X}{L^{w + 3}}$.
Based on (\ref{snrVLC}) and (\ref{pdf_In}),  the PDF of the end user instantaneous SNR  is derived as 
\begin{equation}
f_{\gamma_{\rm vlc}}\left( \gamma  \right) = \frac{\left( \mu _{\rm vlc}\mathcal X ^2 \right)^{  \frac{1}{w + 3}}}{r_{w}^2\left( w + 3 \right)}\gamma^{ - \frac{w + 4}{w + 3}},\quad \gamma_{\min }\leq \gamma \leq \gamma_{\max },
\label{pdfVLC}
\end{equation}
where $\gamma_{\min} = \frac{\mu_{\rm vlc}  \mathcal X^2}{{\left( r_{w}^2 + L^2 \right)}^{w + 3}}$ and  $\gamma_{\max} = \frac{\mu_{\rm vlc} \mathcal X^2}{L^{2 \left(w + 3\right)}}$, and
\begin{equation}
\mu_{\rm vlc}=\frac{P_t^2\eta ^2}{\sigma _d^2}.
\label{avSNR_VLC}
\end{equation}
The CDF of the instantaneous SNR of the end user is 
\begin{equation}
F_{\gamma_{\rm vlc}}\!\!\left( \gamma  \right)\!\! =\!\! \left\{ {\begin{array}{*{20}{c}}
\!\!\!{1\! +\! \frac{L^2}{r_{w}^2}\! - \!\frac{1}{r_{w}^2}{{\left( {\frac{ \mu _{\rm vlc}\mathcal X ^2  }{\gamma}} \right)}^{\frac{1}{{w + 3}}}},}  \gamma_{\min}\!\!\leq \!\!\gamma \leq\gamma_{\max}\\
{1,}  \quad \quad \quad \quad \quad \quad \quad \quad \quad \quad \gamma > \gamma_{\max}\\
\end{array}} \right.\!\!\!\!.
\label{cdfVLC}
\end{equation}

In used analytical stochastic models, a position of terminal is implicitly contained in the fading/gain distributions. We assume that side effects of mobility, such Doppler frequency shift, are handled by proper receiver design \cite{R2}.

\section{Outage probability  analysis of mixed RF-VLC system}

The outage probability is defined as the probability that the instantaneous end-to-end SNR, $\gamma^{(\rm af)} _{\rm eq}$ or $\gamma^{(\rm df)} _{\rm eq}$, falls below a predetermined outage threshold, $\gamma_{\rm th}$.

\newcounter{MYtempeqncnt}
\begin{figure*}[!t]
\setcounter{MYtempeqncnt}{\value{equation}}
\setcounter{equation}{28}
\begin{equation}
\begin{split} \label{Pout_final}
 \mathcal{P}^{\rm (af)}_{\rm out}&= F^{\rm (af)}_{\rm eq}\!\left(\gamma_{\rm th} \right)= 1- \frac{\left( \mu _{\rm vlc}\mathcal X ^2 \right)^{  \frac{1}{w + 3}}}{r_{w}^2\left( w + 3 \right)}  K\sum\limits_{p = 0}^{K - 1}\sum\limits_{\bigtriangleup = p}  \sum\limits_{i = 0}^B \frac{ A \rho ^i{\left( {1 - \rho } \right)}^{B - i}{\Gamma \left( {m + i} \right){e^{ - Q{\gamma _{\rm th}}}}}}{{{\left( 1 + p\left( {1 - \rho } \right) \right)}^B}{\left( {1 + p} \right)}^{m + i}} \sum\limits_{q = 0}^{m + i - 1} {\sum\limits_{r = 0}^q {{q \choose r}\frac{{{Q^q \gamma _{\rm th}^q {C^r}}}}{{q!}}} }\\
& \times   \left(  {\gamma _{\max }^{-\frac{1}{{w + 3}} - r}}{{{\rm{E}}_{\frac{{w + 2}}{{w + 3}} - r}}\left( {\frac{{QC{\gamma _{\rm th}}}}{{{\gamma _{\max }}}}} \right)} { - {\gamma _{\min }^{-\frac{1}{{w + 3}} - r}}{{{\rm{E}}_{\frac{{w + 2}}{{w + 3}} - r}}\left( {\frac{{QC{\gamma _{\rm th}}}}{{{\gamma _{\min }}}}} \right)}} \right)  
\end{split}
\end{equation}

\vspace*{6pt}

\setcounter{MYtempeqncnt}{\value{equation}}
\setcounter{equation}{39}
\begin{equation}
\begin{split}
 \mathcal{P}_{\rm b}\!& =  \frac{1}{2}- \frac{b^a}{2\Gamma \left( a \right)}\frac{\left( \mu _{\rm vlc}\mathcal X ^2 \right)^{  \frac{1}{w + 3}}}{r_{w}^2\left( w + 3 \right)} K\sum\limits_{p = 0}^{K - 1}\sum\limits_{\bigtriangleup = p}  \sum\limits_{i = 0}^B \frac{ A \rho ^i{\left( {1 - \rho } \right)}^{B - i}{\Gamma \left( {m + i} \right)}}{{{\left( 1 + p\left( {1 - \rho } \right) \right)}^B}{\left( {1 + p} \right)}^{m + i}}  \sum\limits_{q = 0}^{m + i - 1} {\sum\limits_{r = 0}^q {{q \choose r}\frac{{{Q^q  {C^r}}}}{{q!{\left( {b + Q} \right)^{  a + q}}}}} } \\
&\times  \left( {\gamma _{\max }^{ - \frac{1}{{w + 3}} - r}}  \MeijerG*{2}{1}{2}{2}{ 1- a-q ,\,{\frac{{w + 2}}{{w + 3}} - r} }{{ - \frac{1}{{w + 3}} - r},\, 0}{{\frac{{QC}}{{{\gamma _{\max }}\left( {b + Q} \right)}}}}  \right.  - \left.   {\gamma _{\min}^{ - \frac{1}{{w + 3}} - r}}  \MeijerG*{2}{1}{2}{2}{ 1- a-q ,\,{\frac{{w + 2}}{{w + 3}} - r} }{{ - \frac{1}{{w + 3}} - r},\, 0}{{\frac{{QC}}{{{\gamma _{\min }}\left( {b + Q} \right)}}}}  \right)
\label{ber_final}
\end{split}
\end{equation}

\setcounter{equation}{\value{MYtempeqncnt}}
\hrulefill
\vspace*{6pt}
\end{figure*}

\subsection{AF relaying}

If  AF based RF-VLC relaying system is considered, the outage probability can be determined based on (\ref{endSNR}) as the CDF of $\gamma^{\rm (af)}_{\rm eq}$ as 
\setcounter{equation}{24}
\begin{equation}
\begin{split}
\!\!\!\! \mathcal{P}^{\rm (af)}_{\rm out} \!=\! F^{\rm (af)}_{\rm eq}\!\left(\gamma_{\rm th} \right)\!=\! \Pr \!\left( \gamma^{\rm (af)} _{\rm eq} \!<\! \gamma _{\rm th} \right)\!= \!\Pr \!\left( \frac{\gamma_{\rm rf} \gamma_{\rm vlc}}{\gamma_{\rm vlc} + C}< \gamma _{\rm th} \right) 
\label{Pout1}
\end{split}\!,
\end{equation}
where $\Pr  \left( \cdot \right)$ denotes probability. After  conditioning on the instantaneous user SNR, (\ref{Pout1}) is rewritten as
\begin{equation}
\begin{split}
 \mathcal{P}^{\rm (af)}_{\rm out}& = \int\limits_0^\infty  \Pr \left( \gamma_{\rm rf} < \gamma _{\rm th} + \frac{\gamma_{\rm th}  C}{\gamma _{\rm vlc}} \right) f_{\gamma _{\rm vlc}}\left( \gamma _{\rm vlc} \right){\rm d} \gamma_{\rm vlc} \\
& = \int\limits_0^\infty  F_{\gamma _{\rm rf}} \left( \gamma _{\rm th} + \frac{\gamma _{\rm th} C}{x}\right) f_{\gamma _{\rm vlc}}\left( x \right){\rm d} x,
\label{Pout2}
\end{split}
\end{equation}
where  $F_{\gamma _{\rm rf}}  \left( \cdot \right)$ and $f_{\gamma _{\rm vlc}}  \left( \cdot \right)$  are CDF and PDF defined in (\ref{cdf_rf}) and (\ref{pdfVLC}), respectively.

After substituting (\ref{cdf_rf}) and (\ref{pdfVLC}) into (\ref{Pout2}) and utilizing the basic PDF properties $\left( \int\limits_{\gamma _{\min}}^{\gamma _{\max }} f_{\gamma _{\rm vlc}}\left( x \right) {\rm d} x = 1 \right) $, the outage probability is rewritten as
\begin{equation} 
\begin{split}
& \mathcal{P}^{\rm (af)}_{\rm out}  = 1- \frac{\left( \mu _{\rm vlc}\mathcal X ^2 \right)^{  \frac{1}{w + 3}}}{r_{w}^2\left( w + 3 \right)} \\
 &\times K\sum\limits_{p = 0}^{K - 1}\sum\limits_{\bigtriangleup = p}  \sum\limits_{i = 0}^B \frac{ A \rho ^i{\left( {1 - \rho } \right)}^{B - i}}{{{\left( 1 + p\left( {1 - \rho } \right) \right)}^B}{\left( {1 + p} \right)}^{m + i}} \\
& \times \int\limits_{\gamma _{\min}}^{\gamma _{\max }} x^{ - \frac{w + 4}{w + 3}}\Gamma \left( {m + i,Q \left( \gamma _{\rm th}  + \frac{\gamma _{\rm th} C}{x}\right) } \right) {\rm d}x. 
\end{split}
\label{Pout3}
\end{equation}
After utilization of a series representation of  the Incomplete Gamma function by  \cite[(8.352.2)]{grad} and binomial theorem  \cite[(1.111)]{grad}, the outage probability is expressed as
\begin{equation} 
\begin{split}
& \mathcal{P}^{\rm (af)}_{\rm out}  = 1- \frac{\left( \mu _{\rm vlc}\mathcal X ^2 \right)^{  \frac{1}{w + 3}}}{r_{w}^2\left( w + 3 \right)} \\
 &\times K\sum\limits_{p = 0}^{K - 1}\sum\limits_{\bigtriangleup = p}  \sum\limits_{i = 0}^B \frac{ A \rho ^i{\left( {1 - \rho } \right)}^{B - i}{\Gamma \left( {m + i} \right){e^{ - Q{\gamma _{\rm th}}}}}}{{{\left( 1 + p\left( {1 - \rho } \right) \right)}^B}{\left( {1 + p} \right)}^{m + i}} \\
& \times \sum\limits_{q = 0}^{m + i - 1} {\sum\limits_{r = 0}^q {{q \choose r}\frac{{{Q^q \gamma _{\rm th}^q {C^r}}}}{{q!}}} } \int\limits_{{\gamma _{\min }}}^{{\gamma _{\max }}} {\frac{{e^{ - Q\frac{{{\gamma _{\rm th}}C}}{x}}}}{x^{ r + \frac{{w + 4}}{{w + 3}}}}}{\rm{d}}x.
\end{split}
\label{Pout4}
\end{equation}
Integral in (\ref{Pout4}) is solved by utilizing \cite[ (06.34.02.0001.01)]{sajt}. The final closed-form expression for the outage probability of the mixed RF-VLC is 
derived and expressed in (\ref{Pout_final}) on the top of the next
page, where $\rm{E}_\nu \left( \cdot \right) $ denotes the Exponential integral defined in  \cite[(06.34.02.0001.01)]{sajt}.

\textbf{High Average SNR of RF Link Approximation:}
In order to determine the outage probability expression for high average SNR over RF link, the following mathematical manipulations are performed. First, it can be noted that the first term  in constant $ C$  can be neglected for high values of  $\mu_{\rm rf}$.   Hence, based on (\ref{const2}) and (\ref{const}), it holds
\setcounter{equation}{29}
\begin{equation} \label{QC_mi1}
\begin{split}
D&=\left. {\left( {QC} \right)} \right|_{{\mu _{{\rm rf} \to \infty }}} = \\
&=K\sum\limits_{p = 0}^{K - 1} \! \sum\limits_{\bigtriangleup= p }  \sum\limits_{i = 0}^B  \frac{A\rho ^i{\left( {1 - \rho } \right)}^{B - i}{\Gamma \left( {m + i + 1} \right)}}{{{\left( 1 + p\left( {1 - \rho } \right) \right)}^B {{{\left( {1 + p} \right)}^{m + i}}}}}.
\end{split}
\end{equation}
For $\mu_{\rm rf}  \to \infty$, the term $e^{-Q\gamma_{\rm th}} \to 1$  in (\ref{Pout_final}).
 Additionally, the dominant term in the sum over $r$ is the one for $r=q$. After neglecting all terms in (\ref{Pout_final}) except $r=q$, the outage probability floor for $\mu_{\rm rf}  \to \infty$ is derived as
\begin{equation} \label{floor_mi1}
\begin{split}
& \mathcal{P}^{\rm (af)}_{\rm out, \mu_{\rm rf}  \to \infty}= 1- \frac{\left( \mu _{\rm vlc}\mathcal X ^2 \right)^{  \frac{1}{w + 3}}}{r_{w}^2\left( w + 3 \right)}  \\
 &\times K\sum\limits_{p = 0}^{K - 1}\sum\limits_{\bigtriangleup = p}  \sum\limits_{i = 0}^B \frac{ A \rho ^i{\left( {1 - \rho } \right)}^{B - i}{\Gamma \left( {m + i} \right)}}{{{\left( 1 + p\left( {1 - \rho } \right) \right)}^B}{\left( {1 + p} \right)}^{m + i}} \\
&\! \times \sum\limits_{q = 0}^{m + i - 1}  \frac{{{\left(D \gamma _{\rm th}\right)^q }}}{{q!}}\left( \frac{{{{\rm{E}}_{\frac{{w + 2}}{{w + 3}} - q}}\left( {\frac{{D{\gamma _{\rm th}}}}{{{\gamma _{\max }}}}} \right)}}{{\gamma _{\max }^{\frac{1}{{w + 3}} + q}}} { - \frac{{{{\rm{E}}_{\frac{{w + 2}}{{w + 3}} - q}}\left( {\frac{{D{\gamma _{\rm th}}}}{{{\gamma _{\min }}}}} \right)}}{{\gamma _{\min }^{\frac{1}{{w + 3}} + q}}}} \right).
\end{split}
\end{equation}
The outage probability  in (\ref{floor_mi1}) can be utilized to  calculate the outage probability floor when $\mu_{\rm rf}  \to \infty$, which will be demonstrated  in  Section V.

\textbf{High Average LED Power Approximation:}
When the average transmitted LED power is high, i.e., $P_t~ \to \infty$\footnote{Although the results are obtained with assumption of $P_t \to \infty$, they are applicable already for LED powers used in practical systems. More details about utilized realistic transmitted optical power can be found in Section V.} (analogously it holds $\gamma_{\rm max}, \gamma_{\rm min} \to \infty$),  the   values of the arguments of the Exponential integrals  in (\ref{Pout_final}) are sufficiently  small.
Therefore,  \cite[(06.34.06.0029.01)]{sajt} can be applied to perform a series representation of ${\rm{E}}_{\frac{{w + 2}}{{w + 3}} - q}\left(\cdot \right)$, where only the first two terms of performed series representations are taken into account.
Furthermore, the dominant term in the sum over $r$ is the one for $r=0$. After neglecting all summation terms  in (\ref{Pout_final}) except $r=0$, and applying
\begin{equation} 
\sum\limits_{q = 0}^{m + i - 1} {\frac{{{{\left( {Q{\gamma _{\rm th}}} \right)}^q}}}{{q!}}}  = {e^{Q{\gamma _{\rm th}}}}\frac{{\Gamma \left( {m + i,Q{\gamma _{\rm th}}} \right)}}{{\Gamma \left( {m + i} \right)}}
\end{equation}
 based on series representation of the Incomplete Gamma function defined in \cite[(8.352.2)]{grad},  the high LED power  approximation of the mixed RF-VLC is derived as
\begin{equation} \label{Pout_prag}
\begin{split}
\mathcal{P}_{{\rm out}, P_t \to \infty}^{\rm (af)} = F_{\gamma_{\rm rf}}  \left( \gamma_{\rm th} \right),
\end{split}
\end{equation}
where the CDF of the RF link is defined in (\ref{cdf_rf}). Note that obtained approximation is consistent with (\ref{endSNR}) for $\gamma_{\rm vlc} \to \infty$.  It can be concluded that the outage probability performance does not depend on the VLC sub-system conditions when LED power is high.
Using this expression, the outage probability floor can be efficiently calculated.

\textbf{Approximation for $\mu_{\rm rf}\to \infty, P_t  \to \infty$:}
In order to determine the outage probability expression for high average SNR over  RF and high average LED power, i.e., $\mu_{\rm rf}\to \infty,P_t  \to \infty$, we first assume  $P_t \to \infty$, which leads to the expression (\ref{Pout_prag}). Next, we  
set $\mu_{\rm rf}\to \infty$ into (\ref{Pout_prag}), i.e., (\ref{cdf_rf}) and apply  \cite[(06.06.06.0004.02)]{sajt} for Gamma function. After substituting (\ref{const2}) for $Q$ function into obtained expression,  the outage probability approximation is derived as
\begin{equation} \label{app_all}
\begin{split}
\mathcal{P}_{{\rm out}, \mu_{\rm rf}\to \infty, P_t\to \infty}^{\rm (af)}& =  K\sum\limits_{p = 0}^{K - 1}\sum\limits_{\bigtriangleup = p}  \sum\limits_{i = 0}^B \frac{ A \rho ^i{\left( {1 - \rho } \right)}^{B - i}}{{{\left( 1 + p\left( {1 - \rho } \right) \right)}^{B+m+i}}} \\
&\! \times \frac{1}{\left( m + i \right)} \left( \frac{m \gamma_{\rm th}}{\mu _{\rm rf}} \right) ^{m+i}.
\end{split}
\end{equation}
Note that the same expression can be derived by assuming
 $\mu_{\rm vlc}  \to \infty$, i.e., $P_t \to \infty$ into (\ref{floor_mi1}).

\subsection{DF relaying} 

For DF based RF-VLC relaying system,  the outage probability can be determined based on (\ref{endSNRdf}) as the CDF of $\gamma^{\rm (df)}_{\rm eq}$ as 
\begin{equation}
\begin{split}
& \mathcal{P}^{\rm (df)}_{\rm out} = F^{\rm (df)}_{\gamma_{\rm eq}}\left( \gamma_{\rm th} \right) = \Pr \!\left( \gamma^{\rm (df)} _{\rm eq} \!<\! \gamma _{\rm th} \right) =  \\
& = F_{\gamma _{\rm rf}}\left( \gamma_{\rm th} \right) + F_{\gamma _{\rm vlc}}\left( \gamma_{\rm th} \right) - F_{\gamma _{\rm rf}}\left( \gamma_{\rm th} \right)F_{\gamma _{\rm vlc}}\left( \gamma_{\rm th}  \right),
\end{split}
\label{Pout}
\end{equation}
where  the CDFs   $F_{\gamma _{\rm rf}}\left(  \cdot  \right)$ and  $F_{\gamma _{\rm vlc}}\left(  \cdot  \right)$  are previously defined in (\ref{cdf_rf}) and (\ref{cdfVLC}), respectively.

\textbf{High Average SNR of RF Link Approximation:} Due to high average SNR over RF link, it holds $\gamma_{\rm rf} \to \infty$. Applying this to the definition of the instantaneous equivalent end-to-end SNR in (\ref{endSNRdf}), it is obvious that $ \gamma^{\rm (df)}_{{\rm eq}, \gamma_{\rm rf} \to \infty} =  \gamma_{\rm vlc}$. For that reason, the approximation for high average SNR over RF link is defined as
\begin{equation} \label{floor_mi1_df}
\begin{split}
& \mathcal{P}^{\rm (df)}_{\rm out, \mu_{\rm rf}  \to \infty}= F_{\gamma_{\rm vlc}} \left( \gamma_{\rm th} \right),
\end{split}
\end{equation}
where $F_{\gamma_{\rm vlc}} \left( \cdot \right)$ is the CDF defined in (\ref{cdfVLC}). The outage probability approximation does not depend on the RF sub-system conditions when the average SNR is very high.
 
\textbf{High Average LED Power Approximation:} For high average transmitted  LED power, i.e., $P_t  \to \infty$, it is concluded that  $\gamma_{\rm vlc} \to \infty$.  In this case, based on (\ref{endSNRdf}), it holds $ \gamma^{\rm (df)}_{{\rm eq}, P_t \to \infty} =  \gamma_{\rm rf}$.
The outage probability performance for $P_t  \to \infty$ is determined as
\begin{equation} \label{floor_Pt_df}
\begin{split}
& \mathcal{P}^{\rm (df)}_{{\rm out},P_t  \to \infty}= F_{\gamma_{\rm rf}} \left( \gamma_{\rm th} \right),
\end{split}
\end{equation}
where $F_{\gamma_{\rm rf}} \left( \cdot \right)$ is the CDF defined in (\ref{cdf_rf}). Similarly as in previous case, the outage probability approximation for $P_t  \to \infty$ is not dependent on the VLC sub-system conditions.

\textbf{Approximation for $\mu_{\rm rf}\to \infty, P_t \to \infty$:}
Since the approximations for $P_t \to \infty $ for both AF and DF systems, i.e., (\ref{Pout_prag}) and (\ref{floor_Pt_df}), respectively, are the same, based on procedure for AF relaying approximation, it  can be concluded that the approximate outage probability expressions for $\mu_{\rm rf}\to \infty$, $P_t \to \infty$ are also identical for both relaying modes, which is derived and presented  in (\ref{app_all}), thus 
\begin{equation} \label{app_all_df}
\begin{split}
\mathcal{P}_{{\rm out}, { \tiny \mu_{\rm rf} \to \infty,P_t\to \infty} }^{\rm (df)} = \mathcal{P}_{{\rm out}, { \tiny \mu_{\rm rf} \to \infty,P_t\to \infty} }^{\rm (af)}.
\end{split}
\end{equation}

\section{Average BER analysis of mixed RF-VLC system}
 
As another important  metric of the system performance, the average BER expressions are derived  when binary phase-shift keying (BPSK) or differential BPSK (DBPSK) is applied.

\subsection{AF relaying} 

For considered RF-VLC system with AF relaying, the
average BER can be derived based on  \cite[(12)]{ber} as
\begin{equation}
\begin{split}
\mathcal{P}^{\rm (af)}_{\rm b} = \frac{b^a}{2\Gamma \left( a \right)}\int\limits_0^\infty  e^{ - b\gamma }\gamma ^{a - 1}F^{\rm (af)}_{\rm eq} \left( \gamma  \right){\rm d} \gamma 
\label{ber1}
\end{split},
\end{equation}
where  the parameters $a$ and $b$ account for different modulation schemes as $(a, b) = (0.5, 1)$ for BPSK and $(a, b) = (1, 1) $ for DBPSK, and $F^{\rm (af)}_{\rm eq} \left( \cdot \right)$ represents the CDF of the end-to-end SNR determined as the outage probability in (\ref{Pout_final}).

Substituting (\ref{Pout_final}) into (\ref{ber1}), and  following derivation presented in Appendix~\ref{App2}, the average BER expression is derived and expressed in (34) on the top of the previous page,  where $\MeijerG*{m}{n}{p}{q}{ -}{-}{\cdot}$
 represents the  Meijer's \textit{G}-function defined in  \cite[(9.301)]{grad}.
 
\textbf{High Average SNR of RF Link  Approximation:}
The average BER  expression for high average SNR over RF link is derived in a similar way as the outage probability approximation for  $\mu_{\rm rf}  \to \infty$.  Again, the dominant term in the sum over $r$ is the one for $r=q$, thus  all  summation terms in (\ref{ber_final}) except $r=q$ can be neglected. Since for  $\mu_{\rm rf}  \to \infty$ holds $Q \to 0$, thus $(b + Q) \approx  b$. The average BER floor  for $\mu_{\rm rf}  \to \infty$ is derived as
\setcounter{equation}{40}
\begin{equation} \label{BERfloor_mi1}
\begin{split}
& \mathcal{P}^{\rm (af)}_{{\rm b}, \mu_{\rm rf}  \to \infty}=  \frac{1}{2}- \frac{{{b^a}}}{{2\Gamma \left( a \right)}}\frac{\left( \mu _{\rm vlc}\mathcal X ^2 \right)^{  \frac{1}{w + 3}}}{r_{w}^2\left( w + 3 \right)}  \\
 &\times K\sum\limits_{p = 0}^{K - 1}\sum\limits_{\bigtriangleup = p}  \sum\limits_{i = 0}^B \frac{ A \rho ^i{\left( {1 - \rho } \right)}^{B - i}{\Gamma \left( {m + i} \right)}}{{{\left( 1 + p\left( {1 - \rho } \right) \right)}^B}{\left( {1 + p} \right)}^{m + i}} \\
& \times \sum\limits_{q = 0}^{m + i - 1}  \frac{D^q}{{q!b^{a+q}}}  \left( \frac{ \MeijerG*{2}{1}{2}{2}{ 1- a-q ,\,{\frac{{w + 2}}{{w + 3}} - q} }{{ - \frac{1}{{w + 3}} - q},\, 0}{{\frac{D}{{{b\gamma _{\max }}}}}}}{{\gamma _{\max }^{  \frac{1}{{w + 3}} + q}} }  \right. \\
&\! - \left.   \frac{ \MeijerG*{2}{1}{2}{2}{ 1- a-q ,\,{\frac{{w + 2}}{{w + 3}} - q} }{{ - \frac{1}{{w + 3}} - q},\, 0}{{\frac{D}{{{b\gamma _{\min }}}}}}}{{\gamma _{\min }^{  \frac{1}{{w + 3}} + q}} }  \right).
\end{split}
\end{equation}

\textbf{High Average LED Power Approximation:}
Since the high average LED power approximation for the outage probability is determined by (\ref{Pout_prag}) as the CDF of the active RF link, the average BER approximation for $P_t \to \infty$  will  be the average BER of the RF subsystem, i.e., 
\begin{equation}
\begin{split}
\mathcal{P}^{\rm (af)}_{{\rm b}, P_t \to \infty} = \mathcal{P}_{\rm b, rf}.
\label{Pb_prag}
\end{split}
\end{equation}
The average BER of the RF link can derived based on (\ref{ber1}) as
\begin{equation}
\begin{split}
\mathcal{P}_{\rm b, rf} = \frac{b^a}{2\Gamma \left( a \right)}\int\limits_0^\infty  e^{ - b\gamma }\gamma ^{a - 1}F_{\gamma _{\rm rf}} \left( \gamma  \right){\rm d} \gamma 
\label{berRF}
\end{split},
\end{equation}
where  the CDF   $F_{\gamma _{\rm rf}}\left(  \cdot  \right)$ is  previously defined in (\ref{cdf_rf}).  
After following the procedure described in Appendix \ref{App3}, the average BER of the RF link is obtained as 
\begin{equation}
\begin{split}
\mathcal{P}_{\rm b, rf}& = \frac{1}{2} - \frac{K}{2\Gamma \left( a \right)} \sum\limits_{p = 0}^{K - 1}\sum\limits_{\bigtriangleup = p}  \sum\limits_{i = 0}^B A\rho ^i  \frac{1 }{{\left( 1 + p\left( {1 - \rho } \right) \right)}^B} \\
&\times \frac{{\left( {1 - \rho } \right)}^{B - i}}{{\left( {1 + p} \right)}^{m + i}}  \MeijerG*{2}{1}{2}{2}{ 1- a,\,1 }{0,\,m+i}{\frac{Q}{b}}.
\label{intRF2}
\end{split}
\end{equation}
After substituting (\ref{intRF2}) into (\ref{Pb_prag}), the average BER approximation for $P_t \to \infty$  is determined.
When LED power is very high, the average BER expression is independent  on the VLC channel conditions.
By utilizing  expression in (\ref{Pb_prag}), the average BER floor can be efficiently calculated.

\textbf{Approximation for $\mu_{\rm rf}\to \infty, P_t  \to \infty$:}
After substituting the outage probability approximation  (\ref{app_all}) into (\ref{ber1}), integral can be easily solved   utilizing  \cite[(06.05.02.0001.01)]{sajt}. The average BER approximation is derived as
\begin{equation} \label{app_all_ber}
\begin{split}
\mathcal{P}_{{\rm b}, \mu_{\rm rf}\to \infty, P_t\to \infty}^{\rm (af)}& =  \frac{K}{2 \Gamma(a)} \sum\limits_{p = 0}^{K - 1}\sum\limits_{\bigtriangleup = p}  \sum\limits_{i = 0}^B \frac{ A \rho ^i{\left( {1 - \rho } \right)}^{B - i}}{{{\left( 1 + p\left( {1 - \rho } \right) \right)}^{B+m+i}}} \\
&\! \times \frac{\Gamma (a+m+i) }{ m + i} \left( \frac{m }{b \mu _{\rm rf}} \right) ^{m+i}.
\end{split}
\end{equation}

\vspace{-\baselineskip}\vspace{-\baselineskip}

\subsection{DF relaying} 

Assuming DF based RF-VLC system, the
average BER can be determined  as \cite[(12)]{ber} 
\begin{equation}
\begin{split}
\mathcal{P}^{\rm (df)}_{\rm b} = \mathcal{P}_{\rm b, rf} \left( 1-\mathcal{P}_{\rm b, vlc}  \right) + \mathcal{P}_{\rm b, vlc}  \left( 1  - \mathcal{P}_{\rm b, rf} \right),
\label{ber3}
\end{split}
\end{equation}
where $\mathcal{P}_{\rm b, rf}$  and $\mathcal{P}_{\rm b, vlc}$ denote the average BER of the RF and VLC links, respectively. The average BER of the RF link is previously derived in (\ref{intRF2}), while average BER of the VLC link is defined as 
\begin{equation}
\begin{split}
\mathcal{P}_{\rm b, vlc} = \frac{b^a}{2\Gamma \left( a \right)}\int\limits_0^\infty  e^{ - b\gamma }\gamma ^{a - 1}F_{\rm vlc} \left( \gamma  \right){\rm d} \gamma 
\label{berVLC}
\end{split},
\end{equation}
where  the CDF    $F_{\gamma _{\rm vlc}}\left(  \cdot  \right)$  is previously defined in  (\ref{cdfVLC}).
After  substituting (\ref{cdfVLC}) into (\ref{berVLC}), the average BER of the VLC link is obtained  as
\begin{equation}
\begin{split}
 &\mathcal{P}_{\rm b, vlc}=\frac{b^a}{2\Gamma \left( a \right)}\int\limits_{\gamma_{\max}}^{\infty}  e^{ - b\gamma }\gamma ^{a - 1}{\rm d} \gamma
\\
&\!+\!\frac{b^a}{2\Gamma \left( a \right)} \!\!\!\int\limits_{\gamma_{\min}}^{\gamma_{\max}} \!\!\!e^{ - b\gamma }\gamma ^{a - 1}\!
\left(\!  1\! +\! \frac{L^2}{r_{w}^2}\! - \!\frac{1}{r_{w}^2}{\left( {\frac{ \mu _{\rm vlc}\mathcal X ^2  }{\gamma}} \right)}^{\!\!\frac{1}{{w + 3}}} \!\right) \!\!{\rm d} \gamma.
\end{split}
\label{intVLC1}
\end{equation}
Integrals in (\ref{intVLC1}) can be easily solved by applying  \cite[(8.2.32)]{grad} as
\begin{equation}
\begin{split}
&\mathcal{P}_{\rm b, vlc}  = \frac{\Gamma\left( a,b \gamma_{\max}\right)}{2\Gamma \left( a \right)} + \frac{1+ \frac{L^2}{r_{w}^2}}{2\Gamma \left( a \right)}  \\
& \times \left(\Gamma\left( a,b \gamma_{\min}\right) - \Gamma\left( a,b \gamma_{\max}\right) \right)-\frac{ \left( b \mu _{\rm vlc}\mathcal X ^2 \right)^{\frac{1}{w+3}} }{2\Gamma \left( a \right)r_{w}^2}  \\ & \times \left(\Gamma \left(  a - \frac{1}{w + 3},b  \gamma_{\min}\right)  \!- \! \Gamma \left( a - \frac{1}{w + 3},b \gamma_{\max}\right) \!\right).
\end{split}
\label{PeVLC}
\end{equation}

After substituting (\ref{intRF2}) and (\ref{PeVLC}) into (\ref{ber3}), the closed-form expression for the average BER is obtained.

\textbf{High Average SNR of RF Link and  Approximation:}
As it was mentioned in previous Section,  for $\mu_{\rm rf } \to \infty $ it holds $ \gamma^{\rm (df)}_{{\rm eq}, \gamma_{\rm rf} \to \infty} =  \gamma_{\rm vlc}$, thus the average BER for  $\mu_{\rm rf} \to \infty $ is equal to the average BER of the VLC sub-system defined in (\ref{PeVLC}) as
\begin{equation}
\begin{split}
\mathcal{P}^{\rm (df)}_{{\rm b}, \mu_{\rm rf} \to \infty } = \mathcal{P}_{\rm b, vlc} 
\label{ber3a}
\end{split}.
\end{equation}

\textbf{High Average LED Power Approximation:} When $P_t \to \infty$, it has been concluded that  $ \gamma^{\rm (df)}_{{\rm eq}, P_t \to \infty} =  \gamma_{\rm rf}$. Hence,  the average BER for  $P_t\to \infty $ is equal to the average BER of the RF sub-system defined in (\ref{intRF2}) as
\begin{equation}
\begin{split}
\mathcal{P}^{\rm (df)}_{{\rm b}, P_t \to \infty } = \mathcal{P}_{\rm b, rf} 
\label{ber3b}
\end{split}.
\end{equation}

\textbf{Approximation for $\mu_{\rm rf}\to \infty, P_t  \to \infty$:}
Based on (\ref{app_all_df}),  the average BER approximation for DF relaying system will be the same as the one for  AF based relaying system when  $\mu_{\rm rf}\to \infty, P_t \to \infty$ presented in (\ref{app_all_ber}) as
\begin{equation} \label{app_all_ber-df}
\begin{split}
\mathcal{P}_{{\rm b}, \mu_{\rm rf}\to \infty, P_t\to \infty}^{\rm (df)} = \mathcal{P}_{{\rm b}, \mu_{\rm rf}\to \infty, P_t\to \infty}^{\rm (af)}.
\end{split}
\end{equation}

\section{Numerical  and simulation results}

This section presents numerical results, obtained by using derived
analytical expressions, together
with Monte Carlo simulations.
The following values of the parameters are assumed: the photodetector surface area $\mathcal A=1~{\rm cm}^2$, the responsivity $\Re=0.4~{\rm A}/{\rm W}$, the optical filter  gain $T =1$, the refractive index of lens at a photodetector $ \zeta =1.5$. Furthermore, the  conversion efficiency is $\eta=0.8$, the noise power spectral density takes a value $N_0=10^{-21}~{\rm W}/{\rm Hz}$, and the baseband modulation bandwidth is $W= 20~{\rm MHz}$ \cite{vlc1, Haas}.

\begin{figure}[!t]
\centering
\includegraphics[width=3.5in]{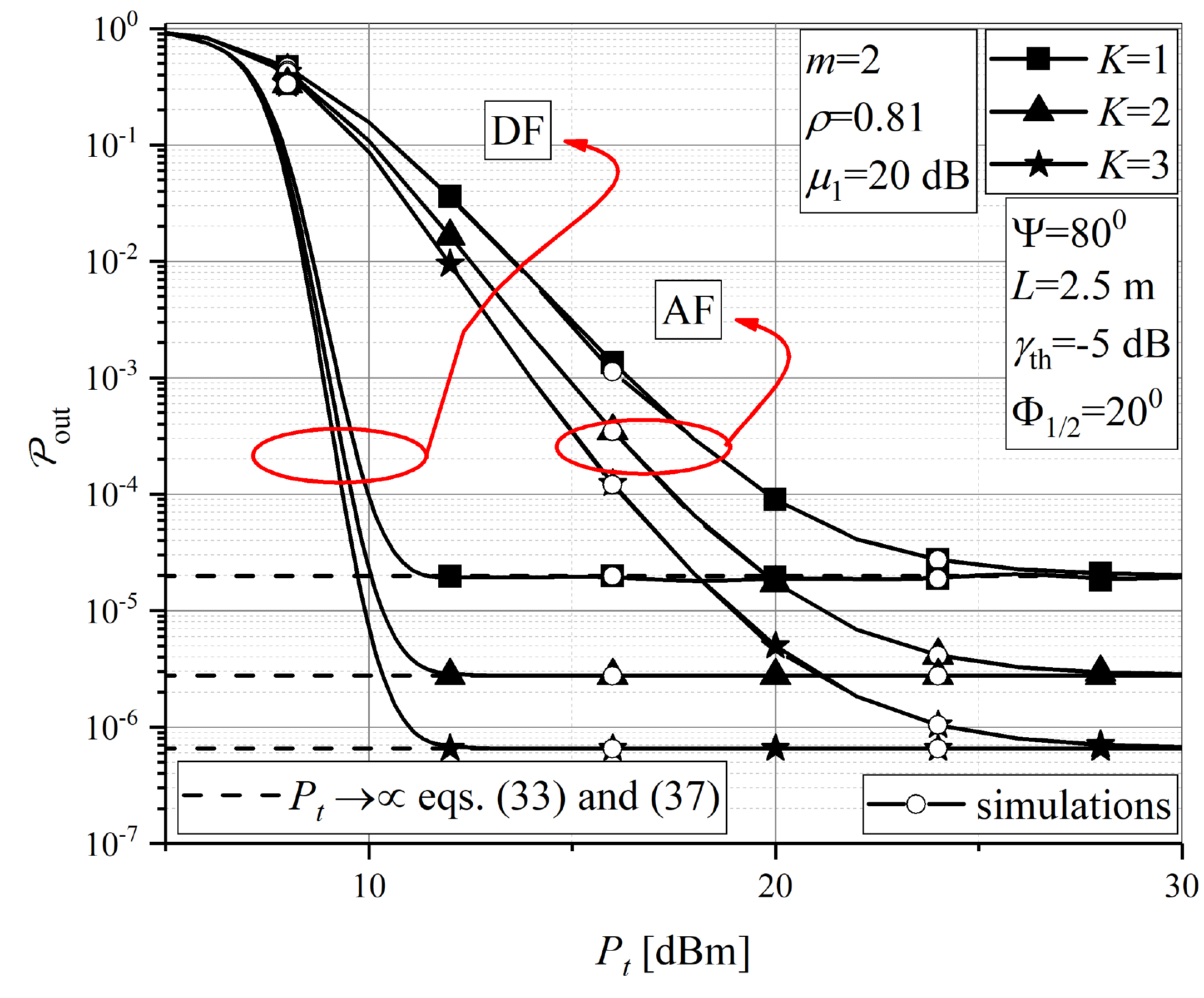}
\caption{Outage probability of the RF-VLC system vs. transmitted optical power.}
\label{Fig3}
\end{figure}

Based on the study presented in \cite{phd}, the following model for LED output power is adopted. Input voltage of LED is 6.42~V, while the  input current  is 700 mA. Hence, the electrical power equals to $P_e = 4.494$ W. Since the  electrical-to-optical conversion efficiency is 0.101, the optical output power of each LED is $P_l = 0.452$ W. In the proposed system, we assume  that the LED lamp consists of  $N_l$  LEDs with the same power $P_l$, i.e.,  $P_t = N_l P_l$ \cite{ISI}.
Depending on the number of the LEDs contained in LED lamp,  the average transmitted optical power of a LED lamp is determined.

\begin{figure}[!t]
\centering
\includegraphics[width=3.5in]{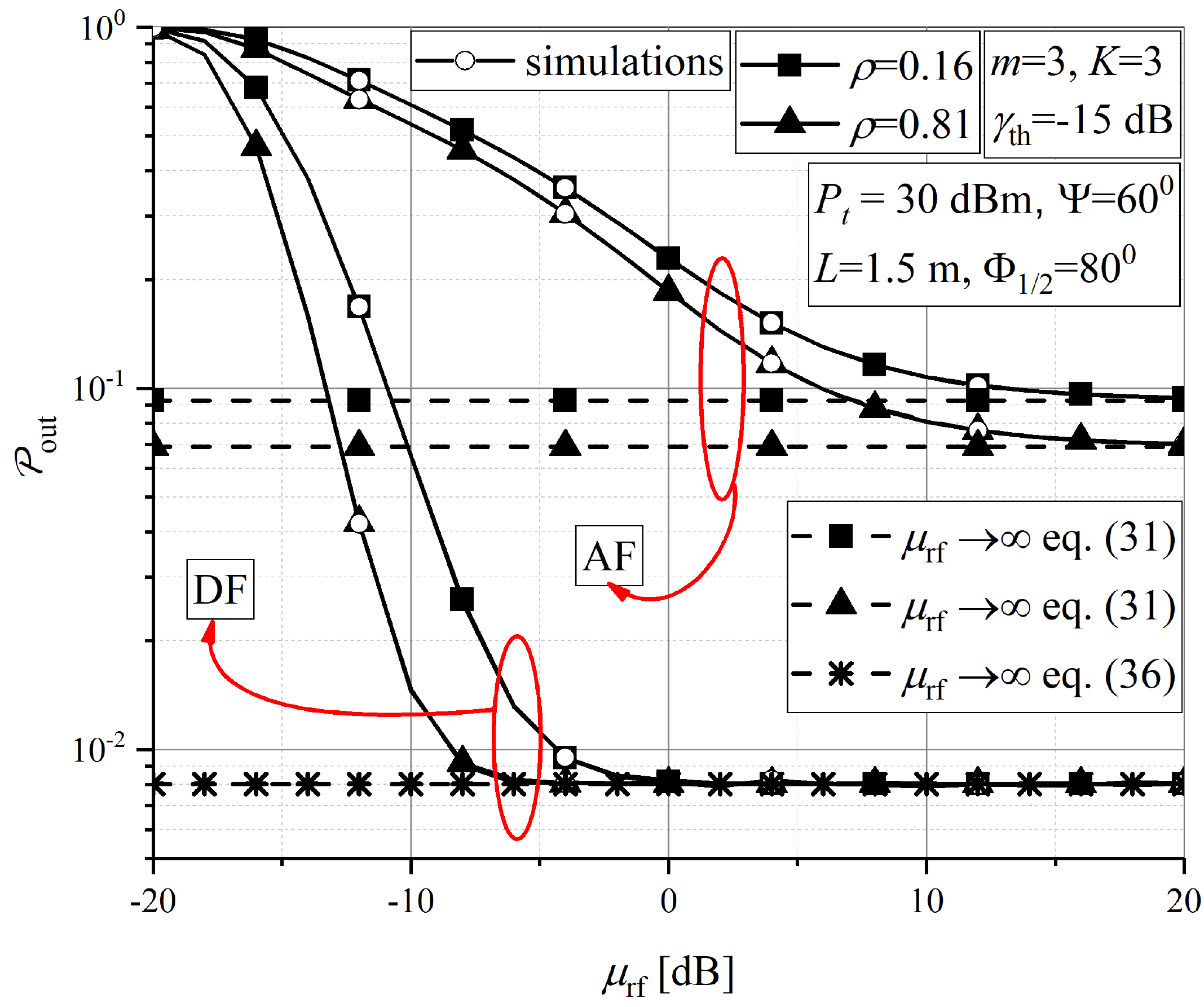}
\caption{Outage probability of the RF-VLC system vs. average SNR of RF link.}
\label{Fig4}
\end{figure}

\begin{figure}[!t]
\centering
\includegraphics[width=3.5in]{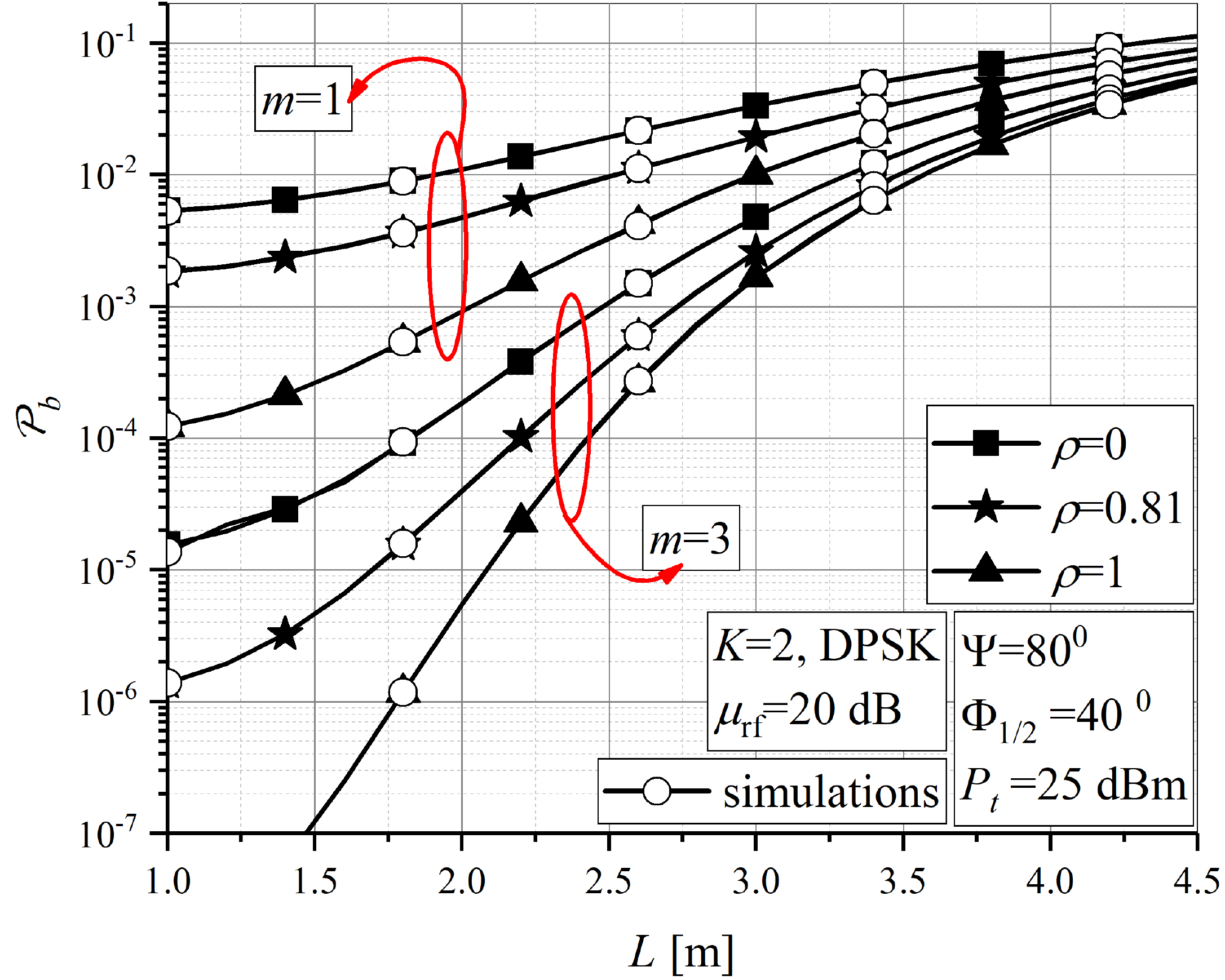}
\caption{Average BER of the RF-VLC system vs. distance between LED lamp and receiving plane.}
\label{Fig5}
\end{figure}

Fig.~\ref{Fig3} shows  the outage probability dependence on the average transmitted optical power of the LED lamp for AF and DF based mixed RF-VLC systems.  
A certain performance gain is noticed when larger number of available BSs is present, which is   independent on the type of implemented relay when $P_t$ is higher.
Furthermore, a certain outage probability floor is noticed for high values of the LED power  in Fig.~\ref{Fig3}. Hence, further increase of the optical signal power will not result in system performance progress. This outage floor appears at lower $P_t$ for DF relaying, but at the same value of $P_t$ for different  number of BSs. 
The outage  floor is in agreement with the derived approximate expressions  (\ref{Pout_prag}) and (\ref{floor_Pt_df})  for AF and DF relaying systems, respectively, which are defined as a CDF  of the instantaneous SNR of the active RF link, $\gamma_{\rm rf}$, defined in (\ref{cdf_rf}). As it can be observed in  Fig.~\ref{Fig3} and confirmed  by  (\ref{cdf_rf}), the outage  floor is determined by the number of BSs related to the RF part of the system. Note that the outage floors appear at the values of $P_t$ which are realistic based on adopted model presented above. For $P_l = 0.452$ W, the LED lamp output power equals to $33.54$ dBm when $N_l=5$ LEDs are employed. Thus, derived expressions for the high average LED power approximation are valid in practical system scenarios.

Outage probability dependence on the average SNR over RF link is depicted in Fig.~\ref{Fig4}, considering both AF and DF relaying. As it is expected, DF relaying system performs better compared to AF one. This justifies choice of fixed gain AF relay only under favorable RF link conditions, i.e., for  higher SNR, corresponding to higher transmission power. Different values of the correlation coefficient  are considered. For lower correlation between  outdated and actual CSIs, the system performance is worsening. In the case of AF relaying, it is noticed that the correlation effect on the system performance is less pronounced  when the $\mu_{\rm rf}$ is lower. When the  outage probability floor occurs, meaning that the further increase in electrical signal power will not improve overall system performance, the impact of the correlation intensity will not be changed with increasing $\mu_{\rm rf}$. The   agreement of  AF outage  floor  with  derived expression  (\ref{floor_mi1}) is observed. The outage floor for  $\mu_{\rm rf} \to \infty$ is dependent on the correlation conditions.  On the other hand, for the case of DF relaying, the outage probability floor for $\mu_{\rm rf} \to \infty$ is  independent on the correlation coefficient. This outage floor is in agreement with derived expression in (\ref{floor_mi1_df}), which is determined to be independent on the RF subsystem conditions. 

To conclude, from Fig.~\ref{Fig3}, as well as from derived expressions  (\ref{Pout_prag}) and (\ref{floor_mi1_df}), it is observed that 
the outage probability floors for great LED power  are equal to the CDF of  the instantaneous SNR of the active RF link, which is independent on the VLC subsystem conditions, for both DF and AF relaying.
From Fig.~\ref{Fig4} and derived expressions (\ref{floor_mi1}) and (\ref{floor_mi1_df}) can be concluded that the outage floors for $\mu_{\rm rf} \to \infty$ is independent on the RF system parameters for DF relaying. On the other hand, the outage floor  for AF relaying  system is dependent on both RF and VLC system parameters. These outage probability floors play important role in determination of system performance, and should be taken into consideration during RF-VLC system design.

In Fig.~\ref{Fig5},  the average BER dependence on the distance between LED and receiving plane  is depicted. Different values of the correlation coefficient are considered. When $L$ is higher, i.e., the optical signal propagation path is longer, the overall received power is reduced and the system performance is deteriorated. From Fig.~\ref{Fig5} it can be concluded that the impact of  $\rho$ on the overall performance is in relation to height $L$. When  distance $L$ is higher, the correlation conditions of the RF link has minor influence on the RF-VLC system performance compared to the case when $L$ is lower. Thus, when the optical receiver is closer to the RF-VLC access point, the RF channel conditions have stronger impact on the overall system performance. Additionally, different values of Nakagami-\textit{m} parameter are considered, describing different fading severities. Greater value of \textit{m} corresponds to decreased fading severity, and system has better overall performance.

\begin{figure}[!t]
\centering
\includegraphics[width=3.5in]{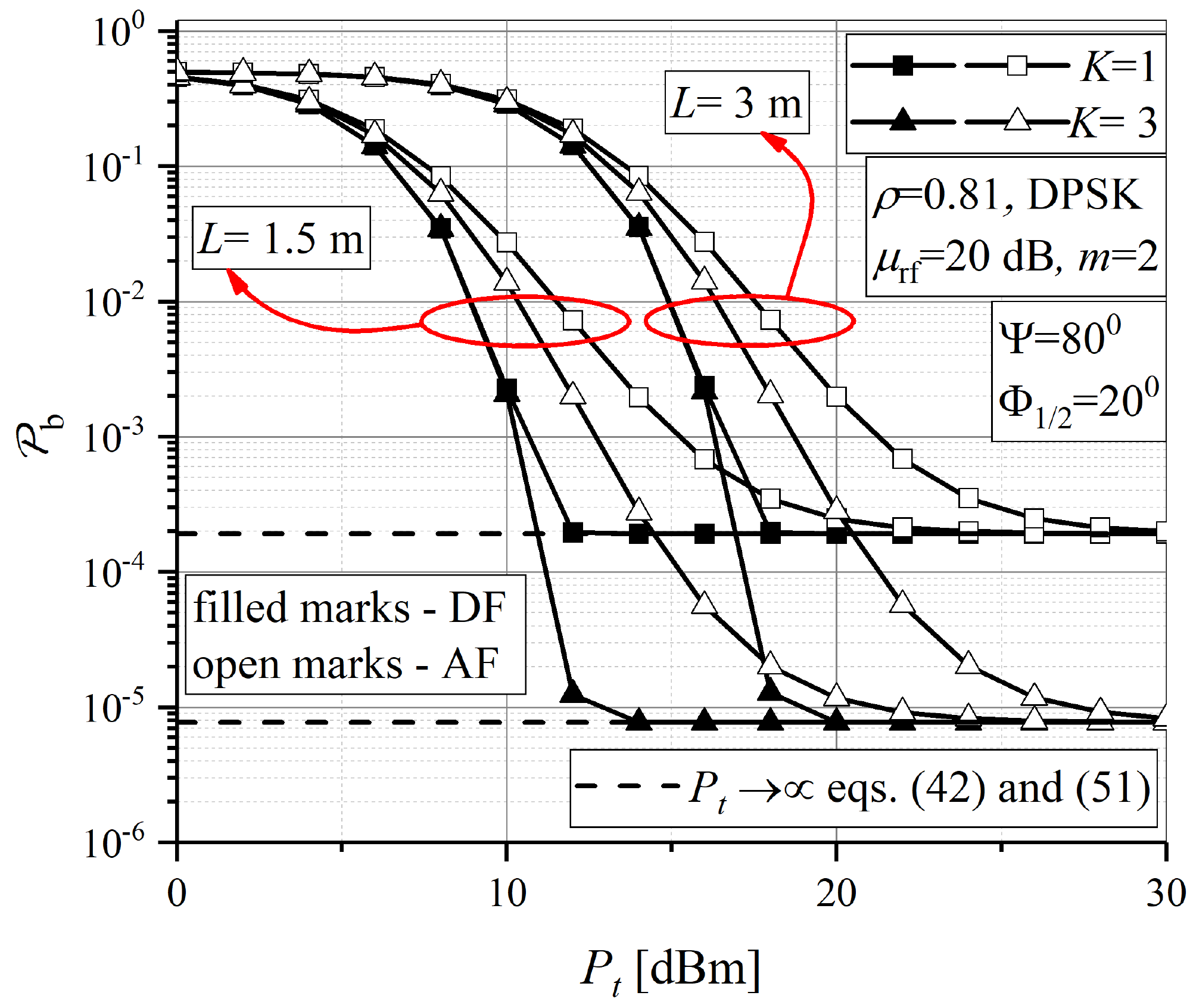}
\caption{Average BER of the RF-VLC system vs.  transmitted optical power.}
\label{Fig6}
\end{figure}

\begin{figure}[!t]
\centering
\includegraphics[width=3.5in]{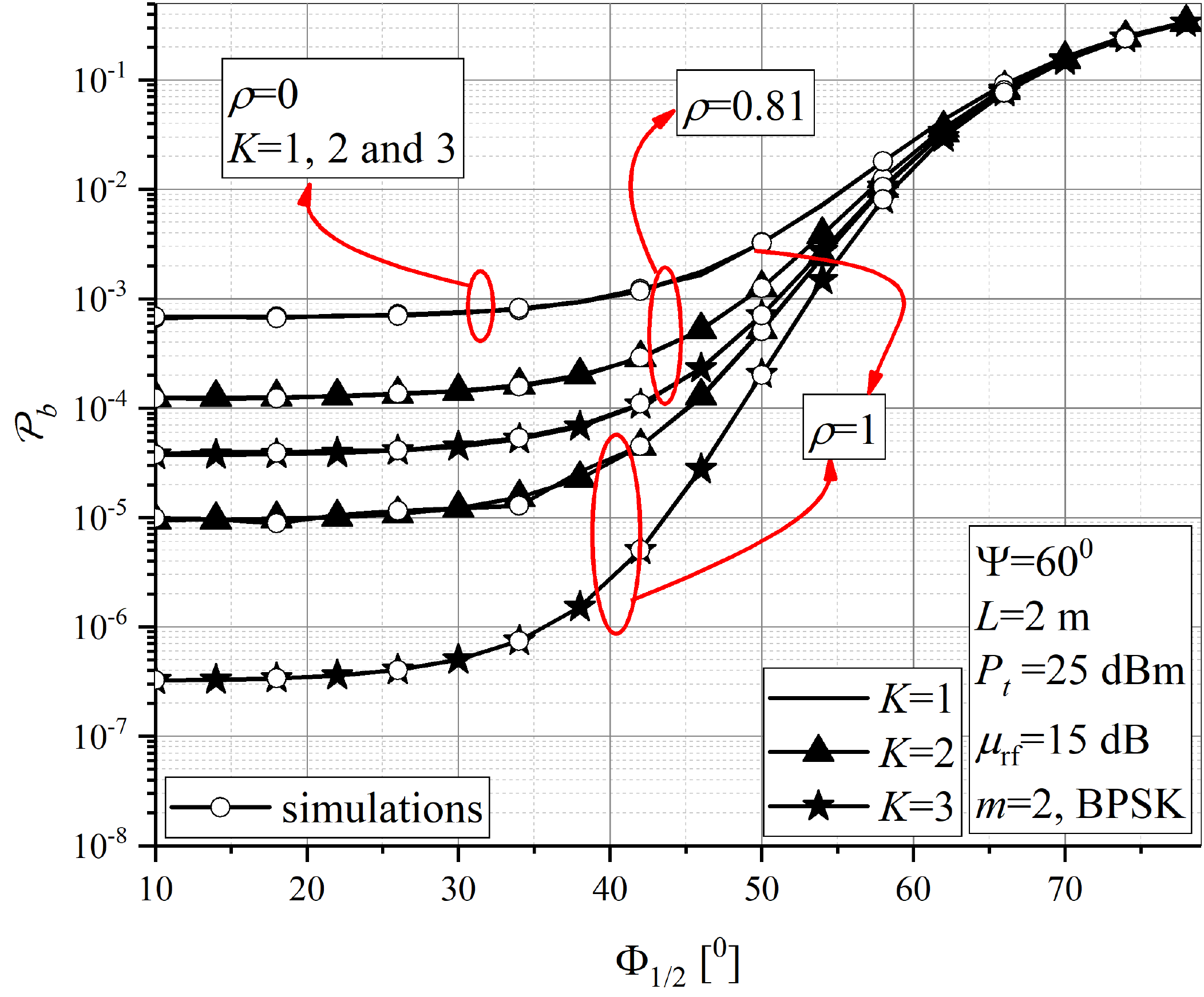}
\caption{Average BER of the RF-VLC system vs. the semi-angle at the half illuminance of LED.}
\label{Fig7}
\end{figure}

In Fig.~\ref{Fig6}  the average BER dependence on the average transmitted optical power of the LED lamp for different heights between LED and receiving plane and various  number of available BSs is shown. Both AF and DF relaying are considered. Greater number of BSs provides a certain performance gain. 
In the range of medium LED power, performance gain is larger when $L$ is lower. However, this difference is lost for high values of $P_t$, since the  floors are independent on $L$.
 To conclude, for lower $P_t$, the improvement due to diversity order is more significant for more favourable VLC  subsystem (lower $\Phi_{1/2}$ and/or lower $L$).

Additionally, the average BER floor is  noticed, which  is  determined by derived expression in (\ref{Pb_prag}) and (\ref{ber3b}) for AF and DF relaying, respectively.  Analogously to the outage floor in (\ref{Pout_prag}) and (\ref{floor_mi1_df}), the BER floor for $P_t$ is independent on  VLC channel conditions (on the distance $L$ in Fig.~\ref{Fig6}), but it is dependent on the RF sub-system, i.e., number of the BSs.

\begin{figure}[!t]
\centering
\includegraphics[width=3.5in]{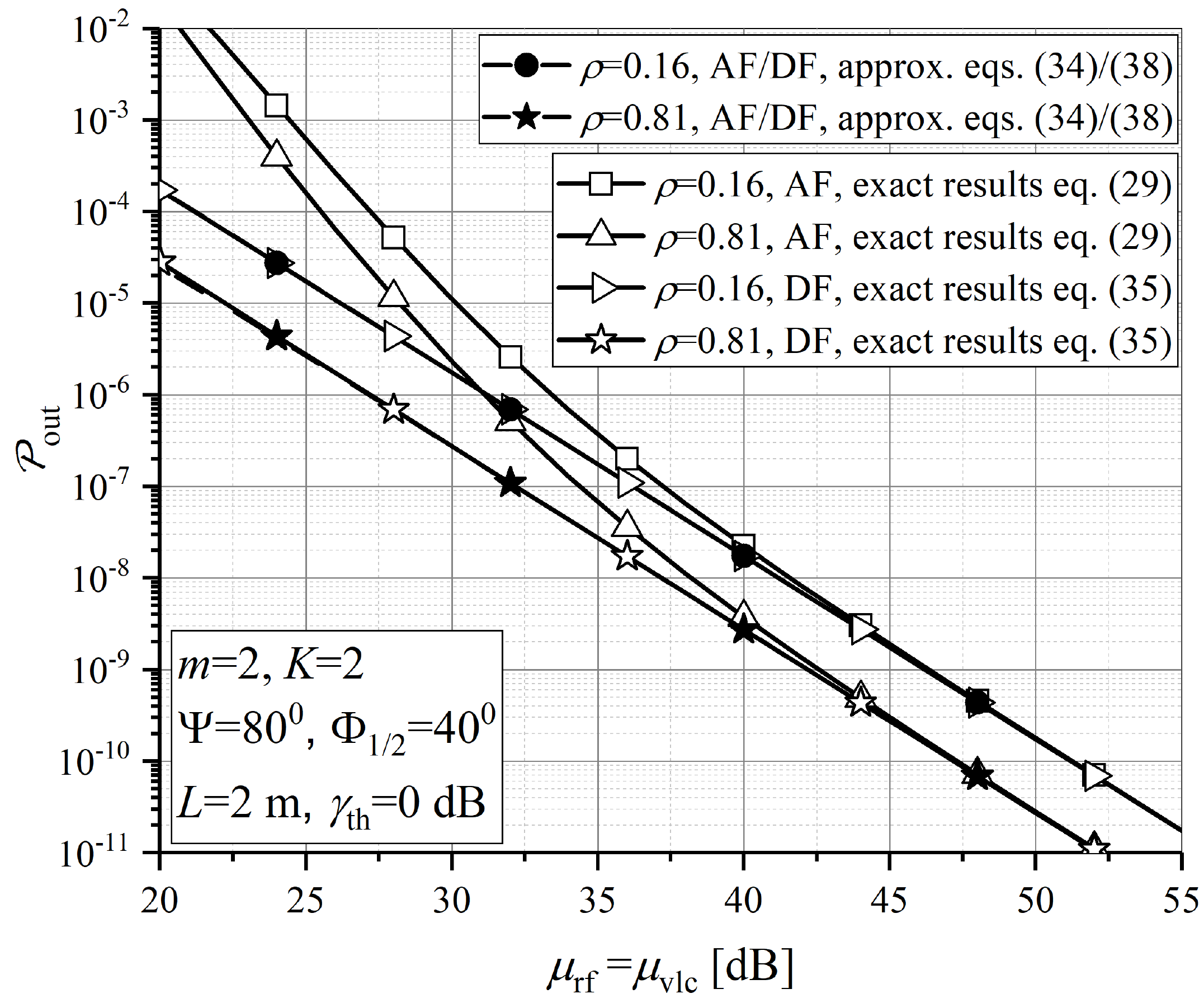}
\caption{Outage probability of the RF-VLC system vs. $\mu_{\rm rf} {\rm [dB]}=P_t  {\rm [dBm]}$ with the corresponding approximations.}
\label{Fig8}
\end{figure}

\begin{figure}[!t]
\centering
\includegraphics[width=3.5in]{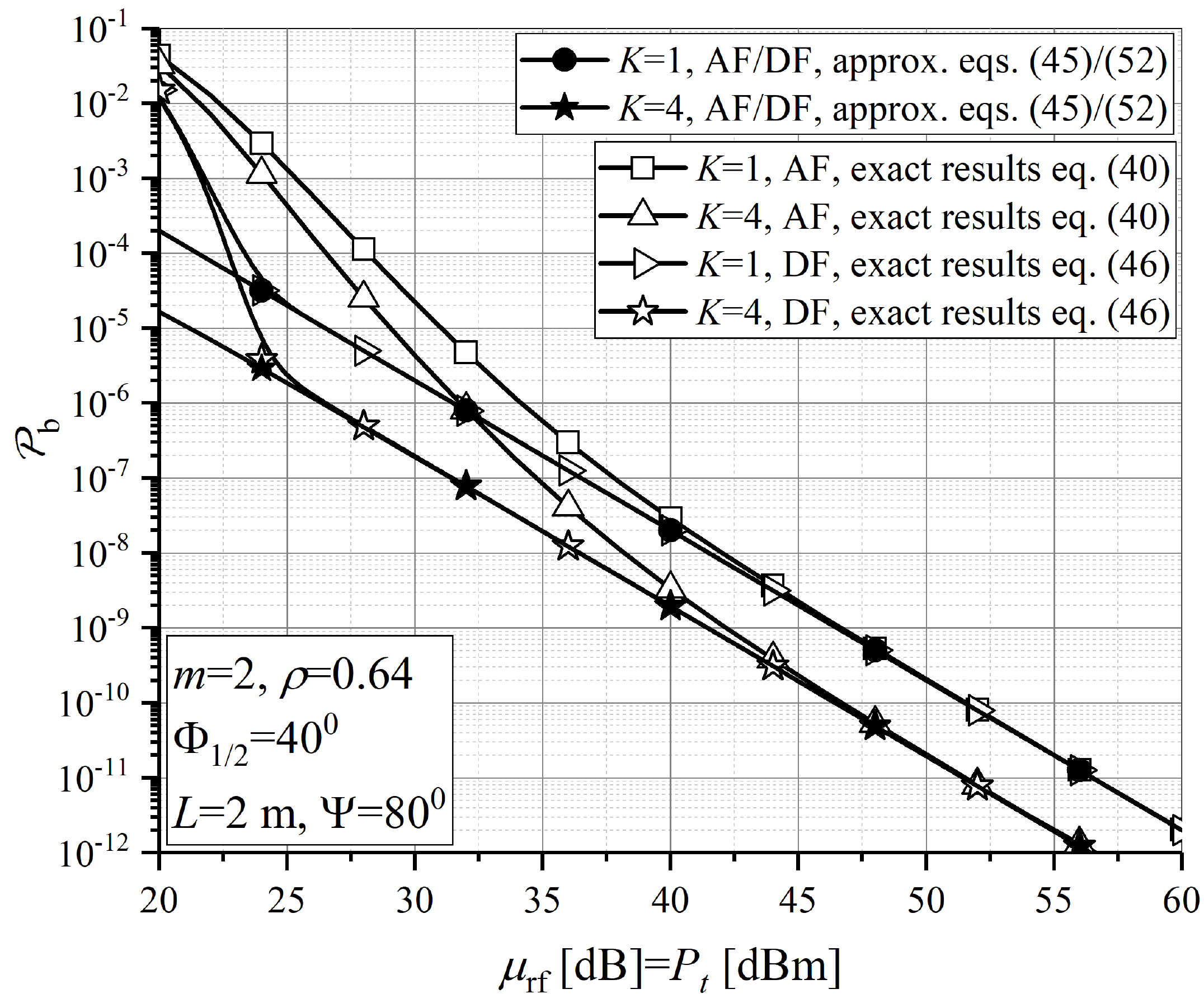}
\caption{Average BER of the RF-VLC system vs. $\mu_{\rm rf} {\rm [dB]}=P_t  {\rm [dBm]}$ with the corresponding approximations.}
\label{Fig9}
\end{figure}

The average BER dependence on the semi-angle at the half illuminance of LED is presented in Fig.~\ref{Fig7}, considering different number of BSs and various correlation coefficient values.  When $\Phi_{1/2}$ is smaller, total received optical power is higher since the optical signal is narrower and more focused, and system performance is better. Contrary, when $\Phi_{1/2}$ is larger, the greater amount of signal energy dissipation exists,  and the total received optical power is reduced, thus performance deterioration exists.
When the semi-angle at the half illuminance of LED lamp is very large, the number of BSs and  correlation coefficient has no influence on the overall system performance. In that case, the energy is distributed  over an excessively large area, thus  very huge energy dissipation exists. Consequently, the RF part of the system will not have important influence on the system performance. Impact of number of BSs is stronger when outdated and actual CSIs are more correlated.  As it is presented in Fig.~\ref{Fig7}, when $\rho =0$, average BER is the same regardless how many BSs are employed. 
In that case, the outdated CSI used for active BS selection and actual CSI are completely uncorrelated, and it can be concluded that  the choice of active BS is insignificant  to the system performance determination.
Also, as CSI becomes more outdated $\left(\rho \searrow \right)$, the impact of  $\Phi_{1/2}$  on average BER is diminishing.

Figs.~\ref{Fig8} and \ref{Fig9} show the outage probability and the average BER dependence on the average SNR over RF link and the LED power, respectively. Approximation for simultaneously  $\mu_{\rm rf} {\rm [dB]} \to \infty $ and $P_t  {\rm [dBm]}\to \infty $  are also presented for both type of relaying modes. It is evident that the derived approximations   are in a very good agreement with the exact expressions in the range of high values of $\mu_{\rm rf} $ and $ P_t$. Furthermore, it can be
observed that approximations become equivalent to  exact results at lower
values of $\mu_{\rm rf}$ and $P_t$ when DF mode is implemented compared
to AF mode.
 Additionally, the accuracy of approximations is independent on the correlation coefficient and the number of available BSs.

\section{Conclusion}

In this paper, we have introduced the statistical analysis of  a mixed  RF-VLC  relaying system suitable for interference-limited mobile applications. Novel
closed-form outage probability and average BER expressions
have been derived considering   radio-access diversity  over mixed RF-VLC system with both fixed gain AF and DF relay. The multiple BSs have been utilized to perform data transmission in outdoor urban environment  by selecting the best link among all RF links based on outdated CSI.
The analytical results have been confirmed via Monte Carlo simulations.

The results have illustrated that the outdated CSI used for active BS selection has an important influence on the end-to-end system performance, especially when the VLC transmission is performed under suitable conditions (lower semi-angle at the half illuminance of LED and lower height between LED lamp and receiving plane).
When the estimated and actual CSIs are uncorrelated, impact of VLC channel and/or number of the available BSs on the overall system performance is minor.
In addition, results have demonstrated that the certain  outage and average BER floors occur at some point. With further increase in  optical or electrical signal power, the system performance improvement will not be accomplished, which is an important limiting factor and should be considered in RF-VLC system design. Based on derived expressions and numerical results, it is concluded that the  floors for great LED power is independent on the VLC sub-system conditions for both AF and DF relaying schemes.
Furthermore, a certain performance  gain has been noticed with employment of multiple BSs. The system performance improvement due to multiple BSs is dependent on the type of employed relaying scheme for medium values of the average LED power, but is independent when the optical power is large and the performance floor exists. Finally, the analysis of ideal end-to-end conditions, for high values of the SNR on RF link $(\mu_{\rm rf} \to \infty)$ and  high LED power $(P_t \to \infty)$, have shown that both outage probability and average BER become equivalent for different (DF and AF) relaying types.

\appendices

\vspace{-\baselineskip}
\section{}
\label{App1}

We assume that RF links experience Nakagami-\textit{m} fading with same fading parameter \textit{m}, thus the PDF and the CDF of instantaneous SNR of each link are given respectively as \cite{Alouni}
\begin{equation} \label{app1-1}
\begin{split}
f_{{\tilde \gamma}_k}\!\left( \gamma \right) \!= \frac{{{m}^m}{\gamma^{m - 1}}}{{{\mu_{\rm rf} ^m}\Gamma \left( m \right)}}e^{- \frac{{m\gamma}}{\mu_{\rm rf} }},~F_{{\tilde \gamma }_k}\!\left( \gamma\right)\! =\!1\! - \!\frac{\Gamma \left( {m,\frac{{m\gamma}}{\mu_{\rm rf} }} \right)}{{\Gamma \left( m \right)}}.
\end{split}
\end{equation}

The active BS is selected according the highest estimated SNR, $\tilde \gamma_n$, which is based on  outdated CSI. 
The PDF of instantaneous SNR of selected BS can be determined
\begin{equation} \label{app1-2}
\begin{split}
f_{{{\tilde \gamma }_n}}\left( \gamma \right) = K{f_{{{\tilde \gamma }_k}}}\left( \gamma \right){\left( {{F_{{{\tilde \gamma }_k}}}\left( \gamma \right)} \right)^{K - 1}}.
\end{split}
\end{equation}
Random variables  $\gamma_k$ and $\tilde \gamma_k$ are correlated with joint PDF given by \cite[(9.296)]{Alouni}
\begin{equation} \label{app1-3}
\begin{split}
{f_{{\gamma _k},{{\tilde \gamma }_k}}}\left( {x,y} \right)& = {\left( {\frac{m}{{{\mu _{\rm rf}}}}} \right)^{m + 1}}\frac{{{x^{\frac{{m - 1}}{2}}}{y^{\frac{{m - 1}}{2}}}}}{{\left( {1 - \rho } \right)\Gamma {{\left( {{m}} \right)} \rho^{\frac{{m - 1}}{2}}}}}\\
& \times {e^{ - \frac{{m \left( {x + y} \right)}}{{\left( {1 - \rho } \right){\mu _{\rm rf}}}}}}{I_{{m} - 1}}\left( {\frac{{2m\sqrt {\rho xy} }}{{\left( {1 - \rho } \right){\mu _{\rm rf}}}}} \right),
\end{split}
\end{equation}
where $I_\nu \left( \cdot \right) $ represents the $\nu$-th order modified Bessel function of the first kind defined in  \cite[(8.406)]{grad}. 

Next, the PDF of the instantaneous SNR of active RF link at the transmission, $\gamma_n$, can be found as
\begin{equation} \label{app1-4}
\begin{split}
{f_{{\gamma _n}}}\left( x \right) = \int\limits_0^\infty  {{f_{{\gamma _n}|{{\tilde \gamma }_n}}}\left( {x|y} \right)} {f_{{{\tilde \gamma }_n}}}\left( y \right){\mathop{\rm d}\nolimits} y,
\end{split}
\end{equation}
where \!\!\!\!\!\!\! 
\begin{equation} \label{app1-5}
\begin{split}
{f_{{\gamma _n}|{{\tilde \gamma }_n}}}\left( {x|y} \right) = \frac{{{f_{{\gamma _k},{{\tilde \gamma }_k}}}\left( {x,y} \right)}}{{{f_{{{\tilde \gamma }_k}}}\left( y \right)}}.
\end{split}
\end{equation}
After substituting (\ref{app1-1}) and (\ref{app1-3}) into (\ref{app1-5}), and afterwards (\ref{app1-2}) and (\ref{app1-5}) into (\ref{app1-4}), the series representation of Gamma function is done by using \cite[(8.352.2)]{grad}. The PDF in  (\ref{app1-4}) is obtained as
\begin{equation} \label{app1-6}
\begin{split}
& f_{\gamma _n}\left( x \right) = K\int\limits_0^\infty  {\left( {\frac{m}{{{\mu _{\rm rf}}}}} \right)}^{m + 1}\!\!\!\!\!\!\frac{x^{\frac{{m - 1}}{2}}{y^{\frac{m - 1}{2}}}}{\left( 1 - \rho  \right)\Gamma \left( m \right)\rho ^{\frac{m - 1}{2}}}e^{ - \frac{{m\left( {x + y} \right)}}{{\left( {1 - \rho } \right){\mu _{\rm rf}}}}}\\
& \!\!\times {I_{{m} - 1}}\!\!\left( {\frac{{2m\sqrt {\rho xy} }}{{\left( {1 - \rho } \right){\mu _{\rm rf}}}}} \right) \!\!{\left( \!\!{1\! -\! {e^{ - \frac{{my}}{{{\mu _{\rm rf}}}}}}\!\sum\limits_{r = 0}^{m - 1} {\frac{1}{{r!}}} {{\left( {\frac{{my}}{{{\mu _{\rm rf}}}}} \right)}^r}} \right)^{\!\!K - 1}}\!\!\!\!\!\!{\mathop{\rm d}\nolimits} y.
\end{split}
\end{equation}
After utilization of  binomial \cite[(1.111)]{grad} and  multinomial theorems, Bessel function is transformed  into Hypergeometric function based on \cite[(03.02.26.0001.01)]{sajt}.  Finally, after performing a series representation of the Hypergeometric function  by \cite[(07.21.07.0002.01)]{sajt}, integral in (\ref{app1-6}) is solved by applying \cite[(07.21.03.0022.01)]{sajt}. The PDF of $\gamma_n$, i.e., $\gamma_{\rm rf}$, is expressed by (\ref{pdf_rf}).
The CDF in (\ref{cdf_rf}) is easily obtained by integrating PDF given in (\ref{pdf_rf}) by utilization  \cite[(8.350.1) and (8.356.3)]{grad}.

\section{}
\label{App2} 

After substituting (\ref{Pout_final}) into (\ref{ber1}), the average BER expression is rewritten as
\begin{equation} \label{app1}
\begin{split}
 \mathcal{P}^{\rm (af)}_{\rm b} & =\Im_1 -  \frac{b^a}{2\Gamma \left( a \right)} \frac{\left( \mu _{\rm vlc}\mathcal X ^2 \right)^{  \frac{1}{w + 3}}}{r_{w}^2\left( w + 3 \right)} \\
&\times N\sum\limits_{p = 0}^{N - 1}\sum\limits_{\bigtriangleup= p }  \sum\limits_{i = 0}^B \frac{ A \rho ^i{\left( {1 - \rho } \right)}^{B - i}{\Gamma \left( {m + i} \right)}}{{{\left( 1 + p\left( {1 - \rho } \right) \right)}^B}{\left( {1 + p} \right)}^{m + i}} \\
& \times \sum\limits_{q = 0}^{m + i - 1} {\sum\limits_{r = 0}^q {{q \choose r}\frac{{{Q^q  {C^r}}}}{{q!}}} }    \left( \Im_2 -\Im_3 \right).
\end{split}
\end{equation}
The first integral $\Im_1$ in (\ref{app1}) is  defined and solved  with the help of \cite[(3.351.3)]{grad} as
\begin{equation} \label{app2}
\begin{split}
\Im_1 = \frac{b^a}{2\Gamma \left( a \right)}\int\limits_0^\infty  e^{ - b\gamma }{\gamma ^{a - 1}} {\rm d}\gamma =  \frac{1}{2}
\end{split}.
\end{equation}
Next, the second integral $\Im_2$ in (\ref{app1}) is  defined  
\begin{equation} \label{app3}
\begin{split}
\Im_2 = \int\limits_0^\infty  \gamma^{q + a-1} e^{ - \gamma \left( b+Q \right)} \frac{{\rm{E}}_{\frac{w + 2}{w + 3} - r}\left( \frac{QC\gamma}{\gamma _{\max }} \right)}{\gamma _{\max }^{\frac{1}{w + 3} + r}}  {\rm d} \gamma 
\end{split},
\end{equation}
while the third one is defined  
\begin{equation} \label{app4}
\begin{split}
\Im_3 = \int\limits_0^\infty  \gamma^{q + a-1} e^{ - \gamma \left( b+Q \right)} \frac{{\rm{E}}_{\frac{w + 2}{w + 3} - r}\left( \frac{QC\gamma}{\gamma _{\min }} \right)}{\gamma _{\min }^{\frac{1}{w + 3} + r}}  {\rm d} \gamma 
\end{split}.
\end{equation}

Integral $\Im_2$ is solved by representing the Exponential integral in terms of the Meijer's \textit{G}-function by \cite[(06.34.26.0005.01)]{sajt} as
\begin{equation} \label{app5}
{{\rm E}_{\frac{w + 2}{w + 3} - r}}\left( \frac{QC\gamma }{\gamma _{\max }} \right) = \MeijerG*{2}{0}{1}{2}{\frac{w + 2}{w + 3} - r }{-\frac{1}{w + 3} - r,\, 0}{\frac{QC\gamma }{\gamma _{\max }}}.
\end{equation}
After replacement (\ref{app5}) in (\ref{app3}), integral $\Im_2$  is solved  with the help of \cite[(07.34.21.0088.01)]{sajt} as
\begin{equation} \label{app6}
\begin{split}
\Im_2& =\gamma _{\max }^{-\frac{1}{w + 3} - r} \left( b+Q \right)^{-a-q}\\
&\times  \MeijerG*{2}{1}{2}{2}{ 1- a-q  ,\,\frac{w+2}{w+3}-r }{-\frac{1}{w+3}-r,\, 0}{\frac{QC}{\gamma _{\max }\left( b+Q \right)}}.
\end{split}
\end{equation}

In the same manner, integral $\Im_3$ is derived as 
\begin{equation} \label{app7}
\begin{split}
\Im_3& =\gamma _{\min }^{-\frac{1}{w + 3} - r} \left( b+Q \right)^{-a-q}\\
&\times  \MeijerG*{2}{1}{2}{2}{ 1- a-q  ,\,\frac{w+2}{w+3}-r }{-\frac{1}{w+3}-r,\, 0}{\frac{QC}{\gamma _{\min }\left( b+Q \right)}}.
\end{split}
\end{equation}

Finally, after substituting (\ref{app2}), (\ref{app6}) and (\ref{app7}) into (\ref{app1}), the final closed-form expression for the average BER of the system under investigation is derived in (\ref{ber_final}).

\section{}
\label{App3} 

After substituting (\ref{cdf_rf})  into (\ref{berRF}), the average BER of the RF link is obtained as 
\begin{equation}
\begin{split}
\mathcal{P}_{\rm b, rf}& = \Im_1 - \frac{b^a}{2\Gamma \left( a \right)} K\sum\limits_{p = 0}^{K - 1}\sum\limits_{\bigtriangleup = p}  \sum\limits_{i = 0}^B  A\rho ^i \\
&\times \frac{{\left( {1 - \rho } \right)}^{B - i}}{{\left( 1 + p\left( {1 - \rho } \right) \right)}^B{\left( {1 + p} \right)}^{m + i}}  \Im_4,
\label{app8}
\end{split}
\end{equation}
where $\Im_1 =  \frac{1}{2}$ has been already defined and solved in (\ref{app2}), while integral $\Im_4$ is defined as
\begin{equation}
\begin{split}
\Im_4 = \int\limits_0^\infty  e^{ - b\gamma }\gamma ^{a - 1}\Gamma \left( {m + i,Q\gamma } \right){\rm d} \gamma.
\label{app9}
\end{split}
\end{equation}
The exponential function in previous integral is represented 
in terms of the Meijer’s \textit{G}-function by using  \cite[(01.03.26.0004.01)]{sajt} as
\begin{equation} \label{app10}
\begin{split}
e^{ - b\gamma} = \MeijerG*{1}{0}{0}{1}{ - }{0}{ b \gamma} 
\end{split},
\end{equation}
while the   Incomplete Gamma function is represented in terms of the Meijer’s \textit{G}-function by \cite[(06.06.26.0005.01)]{sajt} as
\begin{equation} \label{app11}
\begin{split}
 \Gamma \left( m+i,Q\gamma \right) = \MeijerG*{2}{0}{1}{2}{ 1 }{0,\,m+i}{Q\gamma } 
\end{split}.
\end{equation}
After replacement (\ref{app10}) and (\ref{app11}) in (\ref{app9}), integral $\Im_4$  is solved  with the help of \cite[(07.34.21.0011.01)]{sajt} as
\begin{equation} \label{app12}
\begin{split}
\Im_4 =b^{ - a}   \MeijerG*{2}{1}{2}{2}{ 1- a,\,1 }{0,\,m+i}{\frac{Q}{b}}
\end{split}.
\end{equation}
Finally, after substituting (\ref{app2}) and (\ref{app12}) into (\ref{app8}), the final closed-form expression for the average BER of the RF link is derived in (\ref{intRF2}).

\section*{Acknowledgment}

This work has received funding from the European Union Horizon 2020 research and innovation programme under the WIDESPREAD grant agreement No 856967. 


\begin{IEEEbiography}[{\includegraphics[width=1in,height=1.25in,clip,keepaspectratio]{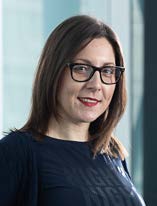}}]{Milica I. Petkovic}
(S’12, M’18)  was born in Knjazevac, Serbia, in 1986. She received her M.Sc. and Ph.D. degrees in electrical engineering from the Faculty of Electronic Engineering, University of Nis, Serbia, in 2010, and 2016, respectively. From 2011 through 2017, she worked as Research Assistant at the Department of Telecommunications, University of Nis. Currently, she is a Post-Doctoral Research Associate at Faculty of Technical Science, University of Novi Sad, Serbia.

Her research interests include optical wireless communications,  wireless communications, cooperative communication,  application of different modulation techniques and modeling of fading channels. 
\end{IEEEbiography}

\begin{IEEEbiography}[{\includegraphics[width=1in,height=1.25in,clip,keepaspectratio]{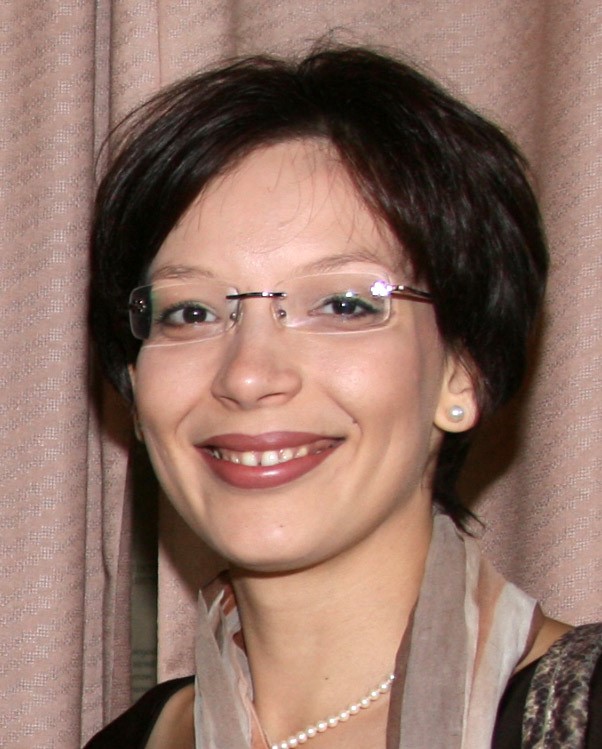}}]{Aleksandra M. Cvetkovic}
 received her BS, MS and PhD degrees in electrical engineering from the Faculty of Electronic Engineering, University of Nis, Serbia, in 2001, 2007 and 2013, respectively. From 2001 to 2019, she worked at the Department of Telecommunications, Faculty of Electronic Engineering, University of Nis. She currently holds the position of Assistant Professor at the Department of Mechatronics and Control, Faculty of Mechanical Engineering, University of Nis.

Her  main research interests include wireless communication systems, communication theory, cooperative communications, wireless power transfer and optical wireless communication systems. 
\end{IEEEbiography}

\begin{IEEEbiography}[{\includegraphics[width=1in,height=1.25in,clip,keepaspectratio]{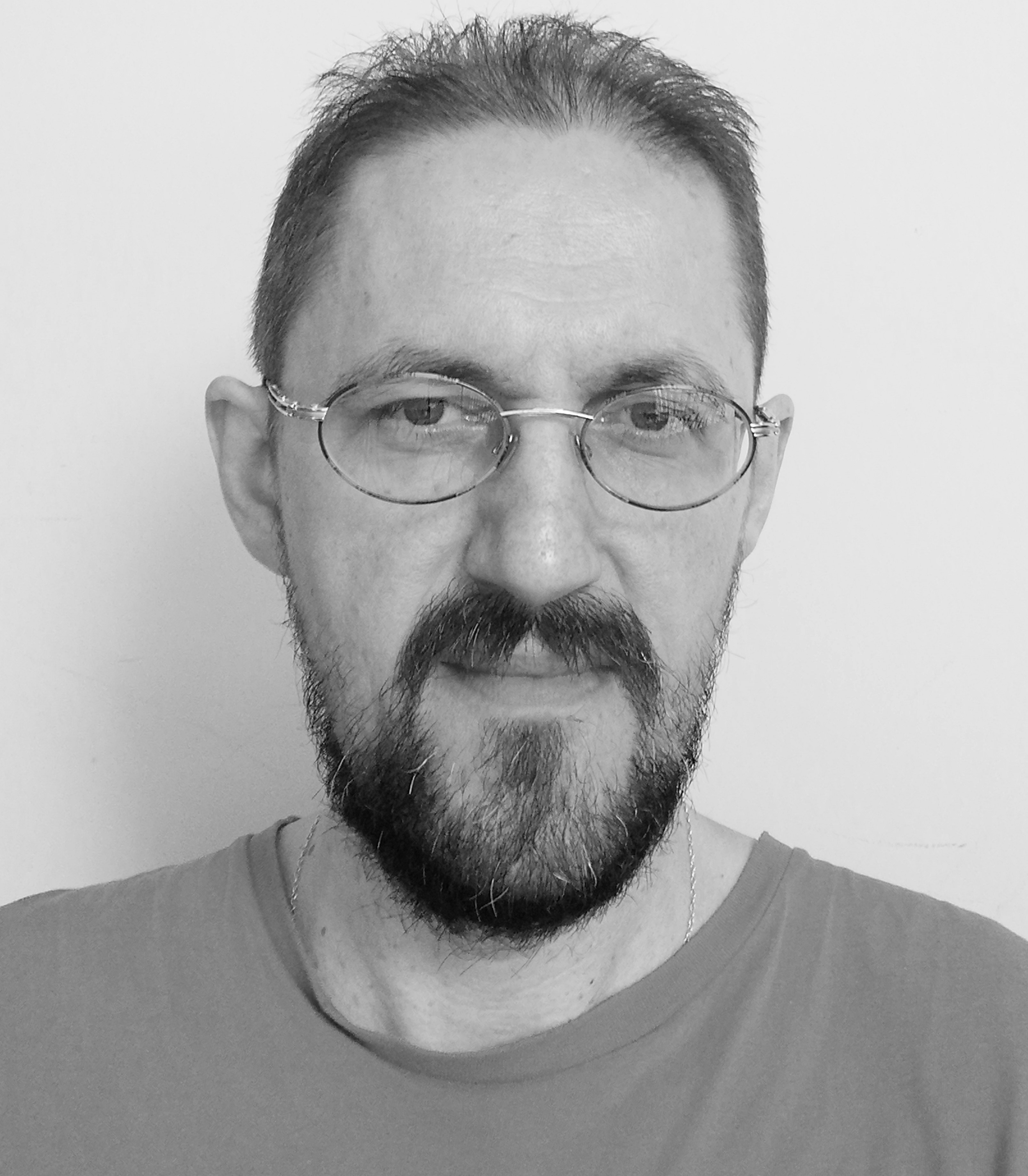}}]{Milan Narandzic}
is Assistant Professor at the Faculty of Technical Sciences, University of Novi Sad. He teaches courses in wireless and mobile communications, cognitive radio, system design, modeling and simulation. He was a Research Assistant at the New Mexico Highlands University, USA (1998-1999) and Technische Universität Ilmenau, Germany (2006-2010), where he received PhD degree in 2015. His research area is related to the spatial aspects of radio/wireless communications (MIMO Channel Modeling, Space-Time Processing, Radio Access Technology …). He participated in many national and international projects. The most important were European FP6 WINNER, CELTIC WINNER+, FP7-ICT EMPhAtiC, ERA.Net HARMONIC and ERASMUS+ BENEFIT. He contributed to many COST (European Cooperation in Science and Technology) actions: 273, 2100, IC1004 and CA15104 IRACON.
\end{IEEEbiography}

\begin{IEEEbiography}[{\includegraphics[width=1in,height=1.25in,clip,keepaspectratio]{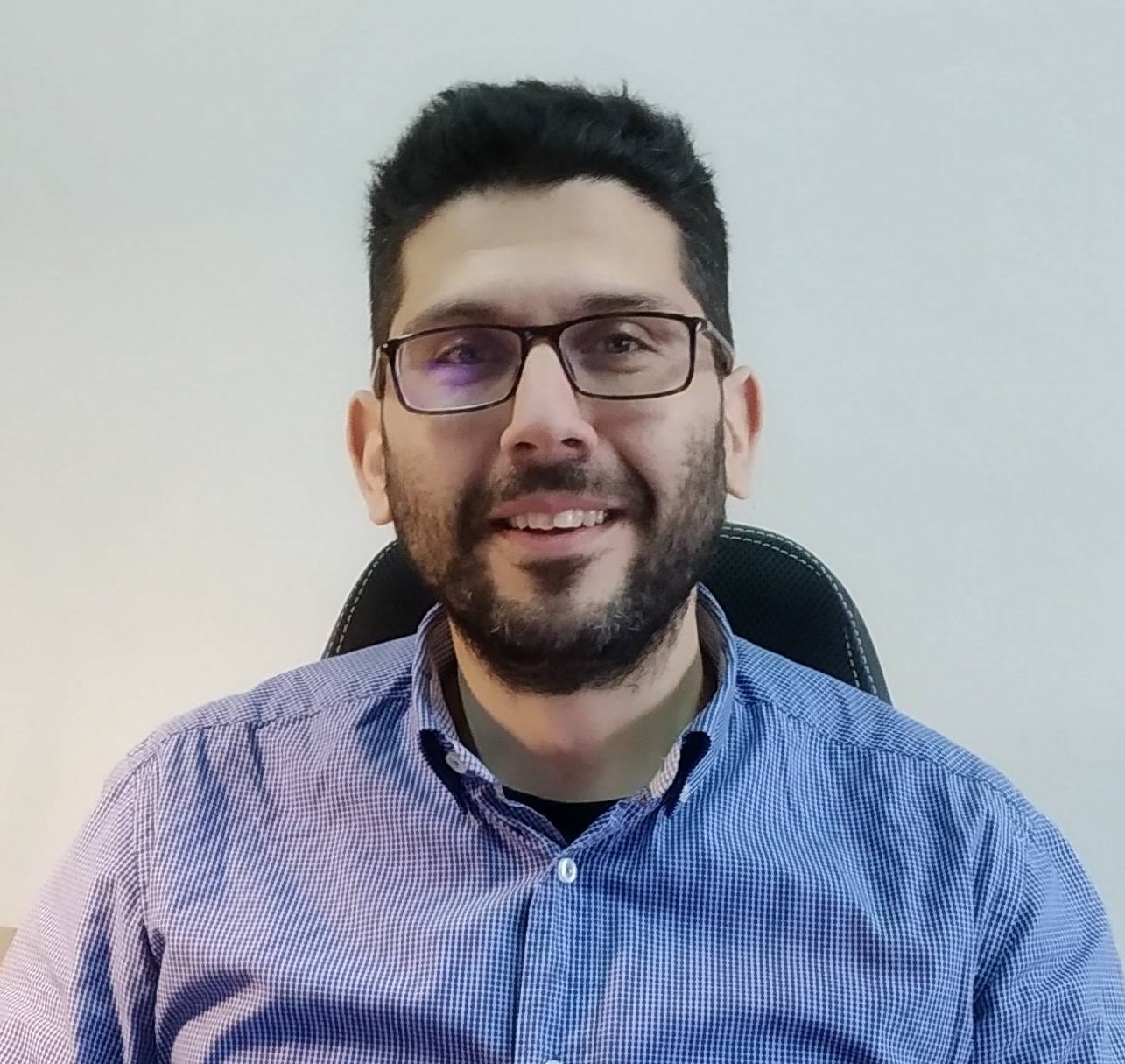}}]{Nestor D. Chatzidiamantis}
(S’08, M’14) was born in Los Angeles, CA, USA, in 1981. He received the Diploma degree (5 years) in electrical and computer engineering (ECE) from the Aristotle University of Thessaloniki (AUTH), Greece, in 2005, the M.Sc. degree in telecommunication networks and software from the University of Surrey, U.K., in 2006, and the Ph.D. degree from the ECE Department, AUTH, in 2012. From 2012 through 2015, he worked as a Post-Doctoral Research Associate in AUTH and from 2016 to 2018, as a Senior Engineer at the Hellenic Electricity Distribution Network Operator (HEDNO). Since 2018, he has been Assistant Professor at the ECE Department of AUTH and member of the Telecommunications Laboratory. 

His research areas span signal processing techniques for communication systems, performance analysis of wireless communication systems over fading channels, communications theory, cognitive radio and free-space optical communications.

\end{IEEEbiography}

\begin{IEEEbiography}[{\includegraphics[width=1in,height=1.25in,clip,keepaspectratio]{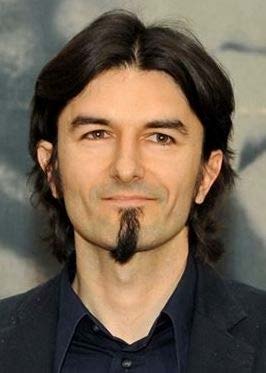}}]{Dejan Vukobratovic}
 (M’09–SM’17) received the Dr.-Ing. degree in electrical
engineering from the University of Novi Sad, Serbia, in 2008. Since
2019. he is a Full Professor 
with the Department of Power, Electronics and Communication Engineering,
University of Novi Sad. From June 2009 until December 2010, he was on
leave as a Marie Curie Intra-European Fellow at the University of Strathclyde,
Glasgow, U.K. From 2011 to 2014, his research was supported by the
Marie Curie European Reintegration Grant. His research group participates
in FP7 ADVANTAGE and H2020 SENSIBLE and H2020 INCOMING EU funded projects. 

His research interests include modern coding theory,
signal processing, probabilistic graphical models and applications in wireless
communication systems. He has
co-authored over 100 research papers mostly published in top-tier IEEE
journals and conferences. 
\end{IEEEbiography}

\begin{IEEEbiography}[{\includegraphics[width=1in,height=1.25in,clip,keepaspectratio]{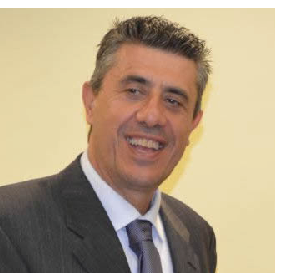}}]{George K. Karagiannidis}
(M’96-SM’03-F’14) was born in Pithagorion, Samos Island, Greece. He received the University Diploma (5 years) and PhD degree, both in electrical and computer engineering from the University of Patras, in 1987 and 1999, respectively. From 2000 to 2004, he was a Senior Researcher at the Institute for Space Applications and Remote Sensing, National Observatory of Athens, Greece. In June 2004, he joined the faculty of Aristotle University of Thessaloniki, Greece where he is currently Professor in the Electrical $\&$ Computer Engineering Dept. and Head of Wireless Communications Systems Group (WCSG).  He is also Honorary Professor at South West Jiaotong University, Chengdu, China.

His research interests are in the broad area of Digital Communications Systems and Signal processing, with emphasis on Wireless Communications, Optical Wireless Communications, Wireless Power Transfer and Applications and Communications $\&$ Signal Processing for Biomedical Engineering.

Dr. Karagiannidis has been involved as General Chair, Technical Program Chair and member of Technical Program Committees in several IEEE and non-IEEE conferences. In the past, he was Editor in several IEEE journals and from 2012 to 2015 he was the Editor-in Chief of IEEE Communications Letters. Currently, he serves as Associate Editor-in Chief of IEEE Open Journal of Communications Society.

Dr. Karagiannidis is one of the highly-cited authors across all areas of Electrical Engineering, recognized from Clarivate Analytics as Web-of-Science Highly-Cited Researcher in the five consecutive years 2015-2019.

\end{IEEEbiography}

\end{document}